\documentclass{article}

\usepackage{microtype}
\usepackage{graphicx}
\usepackage{subfigure}
\usepackage{booktabs} 
\usepackage{soul}

\usepackage{hyperref}
\usepackage[utf8]{inputenc} 
\usepackage[T1]{fontenc}    
\usepackage{url}            
\usepackage{amsfonts}       
\usepackage{nicefrac}       
\usepackage{microtype}      
\usepackage{colortbl}
\usepackage{xcolor}         
\usepackage{graphicx} 
\usepackage{tabularx}
\usepackage{makecell}
\usepackage{longtable}
\usepackage{float}
\usepackage{arydshln}
\usepackage{amsmath}
\usepackage{amssymb}
\usepackage{booktabs}       
\usepackage{xcolor}
\usepackage{dutchcal}
\definecolor{mylightgreen}{rgb}{0.8, 1.0, 0.8}
\definecolor{mylightblue}{rgb}{0.68, 0.85, 0.9}

\definecolor{mycolor}{HTML}{00FF00}

\usepackage[accepted]{icml2024}

\usepackage{amsmath}
\usepackage{amssymb}
\usepackage{mathtools}
\usepackage{amsthm}
\usepackage{siunitx}
\usepackage[capitalize,noabbrev]{cleveref}
\usepackage{enumitem}
\theoremstyle{plain}

\theoremstyle{definition}

\theoremstyle{remark}

\usepackage{longtable}
\usepackage{caption}
\usepackage[title,titletoc]{appendix}

\usepackage{cleveref}

\usepackage[textsize=tiny]{todonotes}

\usepackage{bm} 
\usepackage{xspace} 
\usepackage{multirow} 

\icmltitlerunning{GPT-4V(ision) is a Generalist Web Agent, if Grounded}

\newcommand{\vect}[1]{\ensuremath{\bm{#1}}\xspace}

\newcommand{\lmmweb}{\textsc{SeeAct}}
\newcommand{\gptv}{\textsc{GPT-4V}}

\newcommand{\dataset}{\textsc{Mind2Web}}
\newcommand{\ourmethod}{{\sc {SeeAct}}\xspace}
\newcommand{\eg}{e.g.}

\begin{document}

\twocolumn[
\icmltitle{GPT-4V(ision) is a Generalist Web Agent, if Grounded}



\icmlsetsymbol{equal}{*}

\begin{icmlauthorlist}
\icmlauthor{Boyuan Zheng}{osu}
\icmlauthor{Boyu Gou}{osu}
\icmlauthor{Jihyung Kil}{osu}
\icmlauthor{Huan Sun}{osu}
\icmlauthor{Yu Su}{osu}
\end{icmlauthorlist}

\icmlaffiliation{osu}{The Ohio State University}


\centering{\small \url{https://osu-nlp-group.github.io/SeeAct}}
\icmlkeywords{Machine Learning, ICML}

\vskip 0.3in
]



\printAffiliationsAndNotice{}  

\begin{abstract}
The recent development on large multimodal models (LMMs), especially GPT-4V(ision) and Gemini, has been quickly expanding the capability boundaries of multimodal models beyond traditional tasks like image captioning and visual question answering.
In this work, we explore the potential of LMMs like GPT-4V as a generalist web agent that can follow natural language instructions to complete tasks on any given website.
We propose \ourmethod, a generalist web agent that harnesses the power of LMMs for integrated visual understanding and acting on the web.
We evaluate on the recent \dataset\ benchmark.
In addition to standard offline evaluation on cached websites, we enable a new online evaluation setting by developing a tool that allows running web agents on live websites.
We show that GPT-4V presents a great potential for web agents---it can successfully complete \num{51.1}\% of the tasks on live websites if we manually ground its textual plans into actions on the websites.
This substantially outperforms text-only LLMs like GPT-4 or smaller models (FLAN-T5 and BLIP-2) specifically fine-tuned for web agents.
However, grounding still remains a major challenge. 
Existing LMM grounding strategies like set-of-mark prompting turns out to be not effective for web agents, and the best grounding strategy we develop in this paper leverages both the HTML structure and visuals. 
Yet, there is still a substantial gap with oracle grounding, leaving ample room for further improvement. 
All code, data, and evaluation tools are available at 
{{\small \url{https://github.com/OSU-NLP-Group/SeeAct}}}.
\end{abstract}

\section{Introduction}
\begin{figure}
    \centering
    \resizebox{\linewidth}{!}{
    \includegraphics[width=1\linewidth]{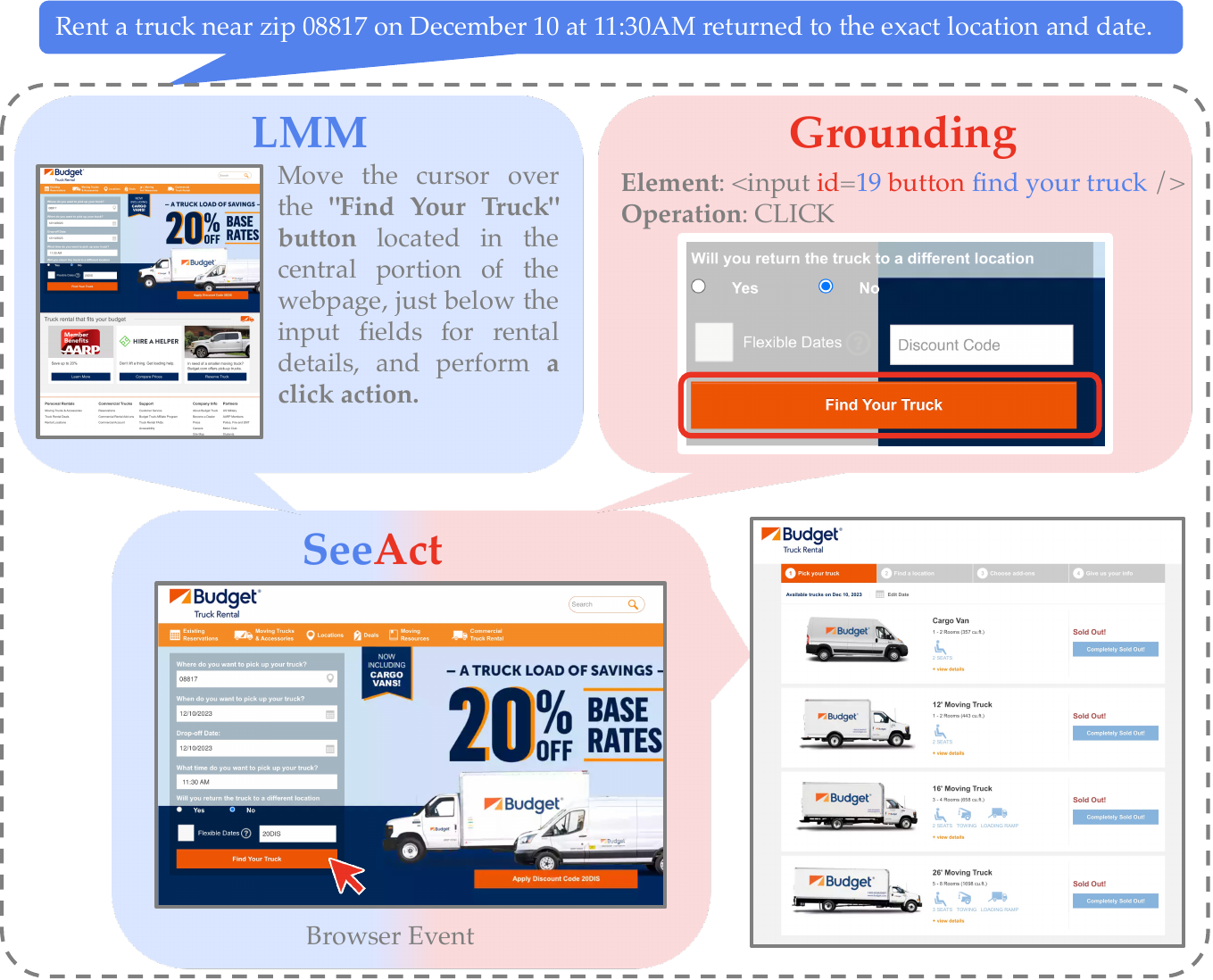}
    }
    \caption{
    \ourmethod\ leverages an LMM like GPT-4V to visually perceive websites and generate plans in textual forms. The textual plans are then grounded onto the HTML elements and operations to act on the website.
    }
    \label{fig:teaser}
\end{figure}

Large multimodal models (LMMs; \citet{Li2023BLIP2BL, Alayrac2022FlamingoAV, Liu2023VisualIT}), especially recent ones such as GPT-4V(ision)~\cite{OpenAI2023GPT4TR} and Gemini~\cite{Anil2023GeminiAF}, have shown a remarkable capability on standard vision-and-language understanding and reasoning benchmarks~\cite{Kazemzadeh2014ReferItGameRT, Goyal2016MakingTV, Hendrycks2020MeasuringMM, Saikh2022ScienceQAAN, Lu2022LearnTE, Zhong2023AGIEvalAH, Yue2023MMMUAM}. While web content has been a primary source of training data, a largely overlooked part of the web is the websites themselves---every website is designed to be rendered visually for easy consumption by human users. 
This poses a new challenge and a new opportunity for LMMs.
On the one hand, screenshots of rendered websites, which could contain thousands of elements with rich relations, are more complex than the images in most existing benchmarks, which are usually object- or scene-centric.
On the other hand, if LMMs can accurately comprehend websites, it will open the door for numerous applications on the web.

In this work, we aim to investigate the potential of LMMs as generalist web agents~\cite{deng2023mind2web}. 
A generalist web agent, as defined in \dataset~\cite{deng2023mind2web}, is expected to follow natural language instructions and complete tasks on any given real-world website (e.g., \autoref{fig:teaser}). 
The tasks can be fairly diverse and complex, with one task possibly taking \num{10}+ actions across multiple dynamically rendered webpages. 
Existing work~\cite{deng2023mind2web,Liu2023AgentBenchEL} primarily uses large language models (LLMs) such as GPT-4~\cite{OpenAI2023GPT4TR} on the raw HTML input. 
However, HTML code is noisier than the rendered visuals and has a lower information density. 
For example, the screenshot in~\autoref{fig:teaser} contains \num{423} HTML elements that would require \num{186490} textual tokens with the GPT-2 Tokenizer, while requiring only \num{1445} visual tokens using GPT-4V's visual tokenizer. 
Furthermore, HTML alone provides incomplete information and misses critical semantics from, e.g., embedded images.

To this end, we propose \ourmethod, a generalist web agent that harnesses the power of LMMs for integrated visual understanding and acting on the web. 
We will focus on GPT-4V, the most advanced LMM publicly available to date, and compare it with smaller LMMs such as BLIP-2~\cite{Li2023BLIP2BL} and LLaVA-1.5~\cite{liu2023improvedllava, liu2023llava}. 
We find that GPT-4V exhibits a strong capability in visually understanding rendered webpages 
and generate the right plans in textual forms across a wide range of websites and tasks.
However, \textit{grounding}~\cite{chandu2021grounding,gu-etal-2023-dont}, \textit{i.e.}, converting the textual plan into precise actions on the website, remains a major challenge. 
It involves selecting the right HTML element to interact with as well as the right operation (e.g., \texttt{Click}, \texttt{Type}, or \texttt{Select}). 
We propose multiple grounding methods, including superpositioning bounding boxes and index labels onto the image, similar to set-of-mark prompting~\cite{Yang2023SetofMarkPU} that has been shown effective on object- or scene-centric images.
However, we find that on complex images with rich semantic and spatial relationships like webpage screenshots, severe hallucination is observed from GPT-4V.
The most effective grounding strategy leverages the known correspondence between HTML element and their visual rendering, a unique property for websites compared to natural images.

We evaluate \ourmethod\ on the \dataset\ dataset~\cite{deng2023mind2web} and compare it with text-only large language models (LLMs) like GPT-4~\cite{OpenAI2023GPT4TR} as well as smaller models (FLAN-T5~\cite{Chung2022ScalingIL}, BLIP-2~\cite{Li2023BLIP2BL}, LLaVA-1.5~\cite{liu2023improvedllava, liu2023llava}, and CogAgent~\cite{Hong2023CogAgentAV}. 
In addition to the standard offline evaluation setting on cached websites, we further establish a new \textit{online evaluation} setting by developing a tool that allows for running web agents on live websites. 
The major findings from our exploration are summarized below:
\begin{itemize}[itemsep=2pt,topsep=0pt,parsep=0pt,partopsep=0pt]
    \item \ourmethod\ with GPT-4V is a strong generalist web agent, if oracle grounding is provided. In online evaluation, it can successfully complete \num{51.1}\% of tasks on different websites, substantially outperforming existing methods like GPT-4 (\num{13.3}\%) or FLAN-T5 (\num{8.9}\%).
    This strongly demonstrates the potential of LMMs like GPT-4V for web agents.
    \item However, grounding is still a major challenge. The best grounding strategy still has a \num{20}-\num{30}\% gap with oracle grounding. Among the various grounding strategies, the best one organically leverages both HTML text and visuals, substantially outperforming image annotation strategies~\cite{Yang2023SetofMarkPU} by up to \num{10}\%.
    \item In-context learning with large models (both LMMs and LLMs) shows better generalization to unseen websites, while supervised fine-tuning still has an edge on websites seen during training.
    \item There is a non-negligible discrepancy between online and offline evaluation because there can often be multiple viable plans for completing the same task. Online evaluation is more indicative of a model's true performance.
\end{itemize}
\section{SeeAct}
\label{sec:background}
\begin{figure*}[h]
    \centering
    \includegraphics[width=.9\linewidth]{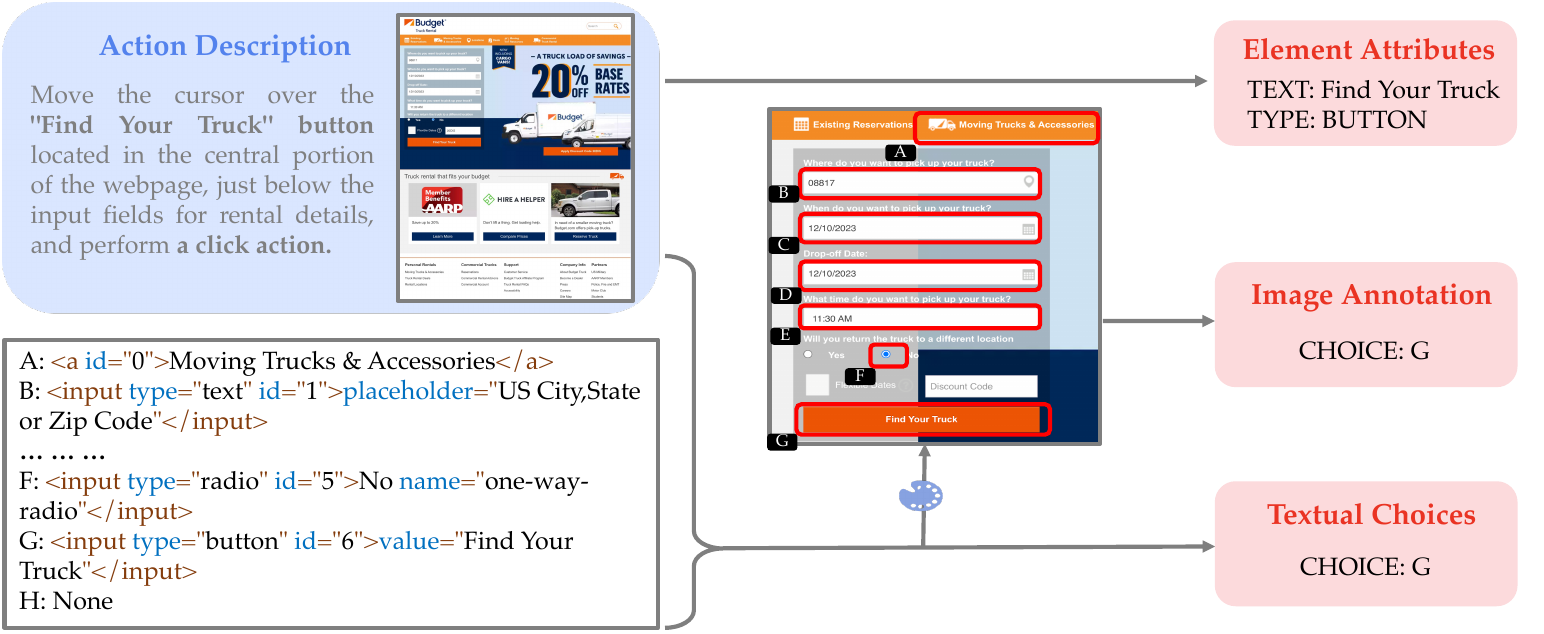}
    \caption{An example of the element grounding process for a single action during completing the given task with three different methods. In this action step, the model needs to click the "Find Your Truck" button to perform a search. 
    For grounding with textual choices, some element candidates represented with HTML text are given, the model is required to generate the choice index of the target element. 
    For image annotation, bounding boxes and index labels are added to the image. The model is required to generate the label on the bottom-left of the target element.  
    For grounding with element attributes, the model needs to predict the text and type of the target element.
    }
\vspace{-10pt}
\label{fig:main-figure}
\end{figure*}

In this section, we first explain the problem formulation of web agents and then introduce \textbf{\ourmethod}, a generalist web agent based on LMMs. Specifically, given a web-based task (\eg, ``Rent a truck with the lowest rate'' in the car rental website), we examine two essential capabilities of LMMs as a generalist web agent: (i) \textbf{Action Generation} to produce an action description at each step (\eg, ``Move the cursor over the `Find Your Truck' button and perform a click'') towards completing the task, and (ii) \textbf{Element Grounding} to identify an HTML element (\eg, ``[button] Find Your Truck'') at the current step on the webpage.

\subsection{Formulation}

Given a website $\mathcal{S}$ (\textit{e.g.,} a car rental website) and a task $T$ (\textit{e.g.,} \emph{``Rent a truck with the lowest rate''}), the web agent should generate a sequence of executable actions $A=[\vect{a}_1, \vect{a}_2,...,\vect{a}_n]$ to complete the task. Specifically, at time step $t$, the agent should generate an action $\vect{a}_t$ based on the current environment observation ${s}_t$, the previous actions $\{\vect{a}_1, \vect{a}_2, ..., \vect{a}_{t-1}\}$, and the task $T$: $$\vect{a}_t = \pi(s_t, T, \{\vect{a}_1, \vect{a}_2, ..., \vect{a}_{t-1}\})$$
The environment observation $s_t$ comprises an HTML document $h_t$ and a screenshot image $i_t$. LLMs can only be grounded on the HTML document, while LMMs can be grounded on both the HTML document and the screenshot image. The website status is updated accordingly after each action: $$s_{t+1} = \mathcal{S}(\vect{a}_t) = \{h_{t+1}, i_{t+1}\} $$ 
For simplicity, in subsequent step-wise formulations, the time step notation $t$ is omitted.

An action $\vect{a}$ corresponds to a browser event provided by the website environment. Therefore, we formulate an action as a triplet of three necessary variables for a browser event $(e, o, v)$. 
$e \in \mathcal{E}$ identifies the target webpage element to operate on, such as the "Find Your Truck" button in \autoref{fig:main-figure}.
$\mathcal{E}$ represents the set of webpage elements within the environment $\mathcal{S}$. The operation $o \in \mathcal{O}$ is the action to be performed on the target element, with $\mathcal{O}$ encompassing all possible operations in $\mathcal{S}$ (\eg, \texttt{Click}, \texttt{Type}). The variable $v$ denotes the additional value needed for a certain operation (\eg, the date \textit{12/10/2023} for a \texttt{Type} operation). 

However, agents based on LLMs or LMMs are typically unable to directly generate the three variables $(e, o, v)$ required for a browser event. Instead, they generate a textual description of the intended action $\tilde{\vect{a}}$, containing information about these variables as $(\tilde{e}, \tilde{o}, \tilde{v})$. This process is referred to as \textbf{Action Generation}. To interact with the environment, a further step is required to convert $\tilde{\vect{a}}$ into $\vect{a}$, which we refer to as \textbf{Action Grounding}.

\subsection{Action Generation}
\label{sec:action-generation}
We explicitly instruct \gptv\ to imitate humans browsing a webpage and analyze the task, webpage, and previous actions. It is asked to generate an action description $\tilde{a}$ based on its analysis and reasoning. We take the screenshot image $i$ as the visual context without utilizing the HTML document $h$ for action generation.

\subsection{Action Grounding}
\label{sec:referring}
Despite the capability of \gptv\ in identifying and describing the next action to complete the given task in natural language, it is still challenging to convert the action description $\tilde{\vect{a}}$ into an executable action $\vect{a}$ within the environment. Deriving operation type $o$ and value $v$ from the action description $\tilde{\vect{a}}$ can be solved through string parsing reasonably well. The key challenge is to identify the target element $e$ from the generated $\tilde{e}$, which we refer to as \textbf{Element Grounding}.

To address this challenge, we explore three approaches using different types of information: Grounding via Element Attributes, Grounding via Textual Choices, and Grounding via Image Annotation, as depicted in Figure~\ref{fig:main-figure}. The prompting details of action generation and grounding are included in  
\autoref{appendix:action-generation}.

\noindent \textbf{Grounding via Element Attributes.}
\label{referring:exp3}
This approach involves prompting the model to generate as detailed attributes of the target element as possible, thereby providing more information to precisely match with the target HTML element. 
Specifically, we prompt the model to not only describe the element $e$, but also specify the target element's type and the textual content in $\tilde{e}$. For example, as illustrated in~\autoref{fig:main-figure}, the model would generate element text as "Find Your Truck" and identify its type as a "BUTTON." Following this, a heuristic search is performed across the DOM elements, using the element text and type to locate matching elements. In cases where a single match is found, it is automatically selected. Otherwise, when multiple matches arise, the model is further prompted to select the final selection.

\noindent \textbf{Grounding via Textual Choices.}
\label{referring:exp4}
The above approach demands precise and sufficient attribute descriptions from \gptv\ and accurate matching by the heuristic search, which can be highly demanding. For instance, many elements may have no textual content or have textual information in a nearby element instead of itself.

Alternatively, we provide the model with textual representations of elements as choices to facilitate grounding, which has already been proven effective in MindAct~\cite{deng2023mind2web}. Specifically, MindAct utilizes a ranking model to select top-$k$ candidate elements $(e_1,e_2, ..., e_k)$ with a pretrained cross-encoder. Each candidate element is represented as a choice in a multi-choice question with its HTML text, as illustrated in~\autoref{fig:main-figure}. After generating the action description $\tilde{\vect{a}}$, the model is further asked a multi-choice question to choose its intended element from the given multiple choices (including a `none' option). 

\noindent \textbf{Grounding via Image Annotation.}
\label{referring:exp2}
Textual representations alone are sometimes insufficient to distinguish similar or identical elements, as illustrated in~\autoref{appendix_case_identical}. Therefore, in this approach, we propose to overlay a bounding box for each candidate element $e$ selected by the ranker
as well as a label around the bounding box\footnote{We use the Supervision library for image annotation: \url{https://supervision.roboflow.com/}} with a label assignment method to avoid overlapping between markups. The model is expected to generate the label corresponding to the target element. 

\noindent \textbf{Oracle Action Grounding.} Ideally, the action description $\tilde{a}$ must encompass all necessary details to precisely identify each variable $(e, o, v)$ of the action triplet. To assess the performance of action generation, an oracle grounding method, which ensures the variables be identified as long as they are mentioned in the action description, is desired. Here we approximate the oracle grounding method by asking human annotators to identify the model's intended actions.

\section{Experiments}

\subsection{Dataset}
\label{sec:dataset}

We evaluate our methods on \dataset~\cite{deng2023mind2web}, a comprehensive dataset encompassing over \num{2000} complex web tasks with annotated actions. This dataset spans \num{137} websites across \num{31} low-level domains, categorized into \num{12} high-level domains. It supports three primary operations: \texttt{Click}, \texttt{Type}, and \texttt{Select}, with \texttt{Hover} and \texttt{Press Enter} operations integrated into \texttt{Click} to avoid ambiguity.

The dataset's test sets aim to measure the generalization of web agents across different tasks, websites, and domains. Specifically, the Cross-Task setting focuses on evaluating agents on tasks that are new to the training data but within included domains and websites. The Cross-Website setting evaluates agents with tasks across \num{10} new websites for each of the top-level domains in the training data. The Cross-Domain setting assesses agent performance on tasks in two top-level domains held out from the training data.

We align each HTML document in the dataset with its corresponding webpage screenshot image from the \dataset\ raw dump, which undergoes human verification to confirm element visibility and correct rendering for action prediction.
This cleaned version of the dataset is called \textit{Multimodal Mind2Web}\footnote{The dataset is released at \url{https://huggingface.co/datasets/osunlp/Multimodal-Mind2Web}.}, with the statistics in \autoref{tab:subset_stat}.

\begin{table*}[t]
\caption{Statistics of the cleaned Multimodal Mind2Web dataset. The average \# of visual tokens is based on OpenAI visual token calculator. 
}
\centering
\small
\begin{tabular}{lccccccc}
\toprule
\multirow{2}{*}{Split} & 
\multirow{2}{*}{\# Tasks} & 
\multirow{2}{*}{\# Domains} &
\multirow{2}{*}{\# Websites} &
\multirow{2}{*}{Avg \#} &
\multirow{2}{*}{Avg \#} &
\multicolumn{2}{c}{Avg \# HTML} \\
\cmidrule(r){7-8}
& &  & & Actions & Visual Tokens & Elements & Tokens \\
\midrule
Train & \num{1009} &\num{17}  &\num{73}  &\num{7.7}  & \num{4240}  &\num{602} &\num{128827} \\

Cross-Domain &\num{694} &\num{13}   &\num{53}  &\num{5.9}  &\num{4314} &\num{494} &\num{91163} \\
Cross-Task &\num{177} &\num{17}   &\num{64}  &\num{7.6}  & \num{4172}  &\num{607} &\num{123274} \\
Cross-Website &\num{142} &\num{9}  &\num{10}  &\num{7.2} &\num{4653} &\num{612} &\num{114358} \\
\bottomrule
\end{tabular}
\vspace{-5pt}
\captionsetup{width=0.85\textwidth} 
\label{tab:subset_stat}
\vspace{-10pt}
\end{table*}

\subsection{Methods}
\label{ref:methods}

\noindent \textbf{SeeAct.}
In grounding via image annotation and textual choices, we first employ the DeBERTa-base cross-encoder from MindAct~\cite{deng2023mind2web} to rank the top 50 elements for better comparison with its text-only counterparts. Then, we cluster elements into groups of \num{17} options for inference. In grounding via element attributes, no candidate element is provided. We experiment all three grounding methods with GPT-4V API, and use the best-performing grounding method for Gemini Pro Vision~\cite{Anil2023GeminiAF}, and LLaVA-1.5~\cite{liu2023improvedllava, liu2023llava}.

\noindent \textbf{MindAct.}
To compare with \ourmethod, we also implement methods based on text-only LLMs and BLIP-2~\cite{Li2023BLIP2BL} following the two-stage strategy of MindAct~\cite{deng2023mind2web}. 
Firstly, we employ the ranker above to pick the top 50 elements. 
Subsequently, the action generation problem is formulated as a multi-choice question answering problem, with the candidate elements as options, including a "None" option if the target element is absent. During inference, elements are clustered into groups of 5 elements, with iterative refinement, until a single choice is made or all options are discarded. We evaluate supervised fine-tuning (SFT) methods using FLAN-T5~\cite{Chung2022ScalingIL} and BLIP-2-T5 and in-context learning (ICL) methods using GPT-3.5 and GPT-4.

\noindent \textbf{Pixel-Level Grounding.}
LMMs can generate target element coordinates in the image via training on datasets augmented with object coordinates, especially for open-sourced models~\cite{Hong2023CogAgentAV, Cheng2024SeeClickHG, You2023FerretRA}. We choose CogAgent~\cite{Hong2023CogAgentAV} as a representative model for this experiment with the ICL method. Details of each method can be found in \autoref{appendix:methods}. 

\subsection{Offline Evaluation}
\label{sec:offline-eval}
We adopt the evaluation metrics utilized in \dataset. \textbf{Element Accuracy} (Ele. Acc) compares the predicted element with the ground-truth elements.
\textbf{Operation F1} (Op. F1) calculates the token-level F1 score for the predicted operation comprised of action and input value. \textbf{Step Success Rate} (Step SR) measures the success of each action step.  A step is successful only if the selected element and the predicted operation are correct. We report macro averages across tasks for these step-wise metrics. \textbf{Success Rate} (SR) measures the success of an entire task. A task is regarded successful only if all steps have succeeded. This metric is stringent without allowing the model any space for exploration and error correction. Therefore, for offline evaluation, we focus on the first three metrics. However, we also conduct online evaluation on live websites for better evaluation on the whole task success rate, as detailed below.

\subsection{Online Evaluation}
We develop a new online evaluation tool using Playwright\footnote{\url{https://playwright.dev/}} to evaluate web agents on live websites (instead of cached websites in offline evaluation). Our tool can efficiently tunnel multimodal inputs from the browser to the agent and convert the predicted action $(e, o, v)$ into a browser event for execution. 
To adhere to ethical standards, our experiments are restricted to non-login tasks in compliance with user agreements, and we closely monitor agent activities during online evaluation to prevent any actions that have potentially harmful impacts, like placing an order or modifying the user profile. 

\sisetup{detect-weight=true, detect-family=true}

\begin{table*}[t]
\caption{Performance of different models. All models under \ourmethod utilize ``Choices'' for grounding. Methods with * mark are conducted on a subset with \num{30} tasks for each task split.
}
\centering
\small
\tabcolsep 3.5pt
\renewcommand\arraystretch{1.0}
\begin{tabular}{llccccccccc}
\toprule
\multirow{2}{*}{} &
\multirow{2}{*}{Model} &
\multicolumn{3}{c}{Cross-Task} &
\multicolumn{3}{c}{Cross-Website} &
\multicolumn{3}{c}{Cross-Domain} \\
\cmidrule(r){3-5} \cmidrule(r){6-8} \cmidrule(r){9-11}
&  & Ele. Acc & Op. F1 & Step SR & Ele. Acc & Op. F1 & Step SR & Ele. Acc & Op. F1 & Step SR \\
\midrule
&\multicolumn{8}{l}{\textbf{Supervised Fine-Tuning}} &\\
\cmidrule(r){1-3}
&\textsc{FLAN-T5-XL}&\bfseries\num{57.1}&\num{75.7}&\bfseries\num{53.5}  
&\bfseries\num{43.8}&\num{67.7}&\bfseries\num{41.1}  
&\bfseries\num{41.4}&\num{65.9}&\bfseries\num{38.9}\\

&\textsc{BLIP-2-T5-XL}&\num{50.1}&\bfseries\num{77.0}&\num{47.0}
&\num{39.4}&\bfseries\num{69.3}&\num{37.0} 
&\num{41.2}&\bfseries\num{69.3}&\num{38.9}\\

\midrule
&\multicolumn{8}{l}{\textbf{In-Context Learning}} &\\
\cmidrule(r){1-3}

&\textsc{GPT-3.5*} &\num{19.4}&\num{59.2}&\num{16.8}  
&\num{14.9}&\num{56.5}&\num{14.1}  
&\num{25.2}&\num{57.9}&\num{24.1}\\

&\textsc{GPT-4*}&\num{40.8}&\num{63.1}&\num{32.3}  
&\num{30.2}&\num{61.0}&\num{27.0}  
&\num{35.4}&\num{61.9}&\num{29.7}\\

\addlinespace[0.1em]\hdashline\addlinespace[0.1em]

&\textsc{CogAgent} &\num{22.4}&\num{53.0}&\num{17.6}  
&\num{18.4}&\num{42.2}&\num{13.4}  
&\num{20.6}&\num{42.0}&\num{15.5}\\

\addlinespace[0.1em]\hdashline\addlinespace[0.1em]

&\lmmweb&& &&& &&& &\\

&\;\; -- \textsc{LLaVA-1.5}&\num{9.7}&\num{65.6}&\num{8.1}  
&\num{9.1}&\num{60.8}&\num{7.5}  
&\num{10.9}&\num{63.9}&\num{8.5}\\

&\;\; -- \textsc{Gemini Pro Vision}&\num{21.5}&\num{67.7}&\num{19.6} 
&\num{17.1}&\num{61.3}&\num{15.4}  
&\num{20.7}&\num{64.3}&\num{18.0}\\

&\;\; -- GPT-4V &\bfseries\num{46.4} &\bfseries\num{73.4} &\bfseries\num{40.2}   &\bfseries\num{38.0}&\bfseries\num{67.8}&\bfseries\num{32.4}  &\bfseries\num{42.4}&\bfseries\num{69.3}&\bfseries\num{36.8}\\

\rowcolor{gray!30} &\;\; -- GPT-4V-Oracle*
&\num{66.4} &\num{79.2}&\num{61.9}  &\num{69.5}&\num{78.9}&\num{65.0}  &\num{72.8}&\num{73.6}&\num{62.1}\\

\bottomrule
\end{tabular}
\vspace{-5pt}
\label{tab:main}
\end{table*}

\section{Results and Analysis}

\subsection{Offline Evaluation Results}

\noindent \textbf{\gptv\ can be a Generalist Web Agent with Oracle Action Grounding.}
Given an effective action grounding method, \gptv\ has the potential to serve as a generalist web agent. Specifically, as described in \autoref{sec:referring}, we provide \gptv\ with an oracle action grounding method (\lmmweb\textsubscript{Oracle}) through human annotation, the model achieves a step success rate of \num{61.9}\%, \num{65.0}\%, and \num{62.1}\% across three test splits, respectively. As shown in \autoref{tab:main}, this method substantially outperforms other models under all metrics across three test splits. 
Specifically, it achieves a \num{8.4}\% step success rate improvement over the second-best method in the Cross-Task setting. 
The performance advantage is more pronounced under the Cross-Website and Cross-Domain settings, where it leads by 23.9\% and 23.2\% step success rates, demonstrating its generality compared with supervised fine-tuning.
This observation is further corroborated within the online evaluation (\autoref{tab:online_evaluation}).

\noindent \textbf{Element Grounding Method Comparison.}
However, there is a noticeable gap between oracle grounding and all three proposed grounding methods, as shown in~\autoref{tab:grounding-oblation}.
\begin{table}[t]
\caption{Step success rate (\%) of GPT-4V on a subset of 30 tasks for each task split with different grounding methods. ``Attributes'', ``Choices'', ``Annotation'', and ``Oracle''  refer to element grounding via Element Attributes, Textual Choices, Image Annotation, and Human Annotation, respectively, as described in \autoref{referring:exp2}. 
}
\centering
\small
\tabcolsep 3.5pt
\renewcommand\arraystretch{1.0}
\begin{tabular}{llccc}
\toprule
\multirow{2}{*}{} &
\multirow{1}{*}{Grounding} &
\multicolumn{1}{c}{Cross-Task} &
\multicolumn{1}{c}{Cross-Website} &
\multicolumn{1}{c}{Cross-Domain} \\
\midrule
&\lmmweb\textsubscript{Attribute}&\num{16.1}   &\num{12.1} &\num{19.0}\\
&\lmmweb\textsubscript{Annotation}&\num{20.3}  &\num{13.9}&\num{23.7}\\
&\lmmweb\textsubscript{Choice} &\num{39.1}  &\num{32.7}  &\num{42.0}\\
\rowcolor{gray!30} &\lmmweb\textsubscript{Oracle}
&\num{61.9} &\num{65.0} &\num{62.1}\\
\bottomrule
\end{tabular}
\vspace{-5pt}
\label{tab:grounding-oblation}
\end{table}
This demonstrates that grounding, especially element grounding, is a major bottleneck. 
Element grounding via textual choice (\lmmweb\textsubscript{Choice}) demonstrates the best performance under all metrics across all settings,  comparable to supervised fine-tuning and showing a substantial improvement over text-only LLMs.

Grounding via image annotation (\lmmweb\textsubscript{Annotation}) offers an intuitive approach and shows promising results in recent work that focuses on object- or scene-centric images~\cite{Yang2023SetofMarkPU}. 
However, we find that on complex images with rich semantic and spatial relationships like webpage screenshots, severe hallucination is observed from GPT-4V. 
Specifically, it often fails to correctly map its generated element description (which is often correct according to oracle grounding) to the right bounding box and index label in the image, leading to a low element accuracy.
This limitation primarily arises from \gptv's weakness in understanding image details and relative spatial location, a topic that we will further delve into in \autoref{sec:hallucination-examples}. We leverage bottom-left number labels around bounding boxes as it is identified as the optimal markups in the ablation study in ~\autoref{appendix:markup-oblation}.

Grounding via element attributes (\lmmweb\textsubscript{Attribute}) also demonstrates inferior performance. This method's effectiveness is primarily limited by its heuristic-based element localization strategy, which depends on textual and locality characteristics. This becomes problematic as not all webpage elements contain text, and sometimes the relevant text is associated with a nearby but distinct element.

\noindent \textbf{LMMs vs. LLMs.}
The \lmmweb\textsubscript{Choice} with GPT-4V demonstrates a substantial performance advantage over the text-only GPT-4 under all three metrics across all three test splits. Specifically, it outperforms GPT-4 in step success rate of \num{6.8}\%, \num{5.7}\%, and \num{12.3}\% on three settings, respectively.
Interestingly, fine-tuned BLIP-2-T5 does not show a noticeable gain over FLAN-T5, despite having additional visual input. 
Several factors may contribute to this. First, the CLIP model used as the image encoder may not be sufficiently adept at image details, as explored by \citet{shen2021much}. This limitation is particularly relevant for our web navigation task, which demands a high level of image detail comprehension. Second, BLIP-2-T5 utilizes an off-the-shelf CLIP model that may not be optimal for webpage screenshots.
Finally, although the screenshots in the test splits are error-free, some of the examples in the training set might contain issues such as rendering failures or inaccuracies when annotators capture the screenshots. 

\begin{table}[h]
\caption{Whole task success rate (\%) under both offline and online evaluation. Offline\textsubscript{0} and Offline\textsubscript{1} refer to no tolerance for error at any step and allowing for error at one step, respectively. 
}
\centering
\small
\tabcolsep 3.5pt
\renewcommand\arraystretch{1.0}
\begin{tabular}{lccc}
\toprule
& Offline\textsubscript{0}& Offline\textsubscript{1} & Online \\
\midrule
FLAN-T5-XL&\num{4.4}&\num{24.4}& \num{8.9}   \\
GPT-4 &\num{1.1}&\num{12.2}&\num{13.3}   \\
\lmmweb\textsubscript{Choice} &\num{3.3}&\num{12.2}& \num{37.8}\\
\lmmweb\textsubscript{Oracle} &\num{13.3}&\num{27.8}&\num{51.1}\\
\bottomrule
\end{tabular}
\vspace{-5pt}
\label{tab:online_evaluation}
\end{table}

\noindent \textbf{SFT vs. ICL.}
We compare SFT and ICL methods to offer insights for developing web agents in different scenarios. ICL (with \ourmethod) demonstrates consistent and robust performance across three test splits. 
ICL is particularly advantageous in scenarios lacking annotations or requiring strong generalization capabilities for new domains and websites. As grounding methods improve towards oracle grounding, ICL is poised to show even stronger performance.
On the other hand, SFT methods show better generalization across tasks on websites already seen during training. 
Considering the high cost of data annotation for web agents and the billions of websites on the Internet, ICL offers a more compelling solution for generalist web agents.
However, if one only needs to develop a strong web agent for a certain website, SFT is still a competitive solution.

\subsection{Online Evaluation Results}
In online evaluation, we pair a web agent with a human annotator, where the human was tasked to monitor agent actions that may change real-world states and determine whether each task was successfully completed. For comparative analysis, we include success rates from offline evaluation, denoted as Offline\textsubscript{0} (allowing zero wrong action) and Offline\textsubscript{1} (allowing one wrong action).
For a fair comparison between offline and online evaluations, we only re-write time-sensitive tasks to ensure they are still valid when the evaluation is conducted. For instance, we update the dates for flight-related tasks. Finally, we conduct the online evaluation on the same subset of total \num{90} tasks from the three test splits. For tasks invalidated due to website change, we resample from the corresponding test split. 

Table~\ref{tab:online_evaluation} shows that the whole task success rate in online evaluation substantially exceeds that of offline evaluation (Offline\textsubscript{0}). This finding suggests that the whole task success rate is likely underestimated in the offline evaluation due to the variability in actions and plans.
In other words, there may be multiple viable plans for a task, but the reference plan in offline evaluation only captures one of them. 

Across all three settings, \lmmweb\textsubscript{Choice} with GPT-4V outperforms both GPT-4 and FLAN-T5-XL by a large margin of over 20\% whole task success rate. Using oracle grounding further improves the performance substantially, reaching a remarkable whole task success rate of \num{51.1}\%.
Although GPT-4 shows much worse performance than FLAN-T5-XL in step success rate under offline evaluation (\autoref{tab:main}), it outperforms FLAN-T5-XL by 4.4\% whole task success rate in the online evaluation.  These results further confirm the potential of large models for generalist web agents compared with fine-tuned small models.

\subsection{Analysis}

\begin{figure}[t]
    \centering
    \includegraphics[width=0.85\linewidth]{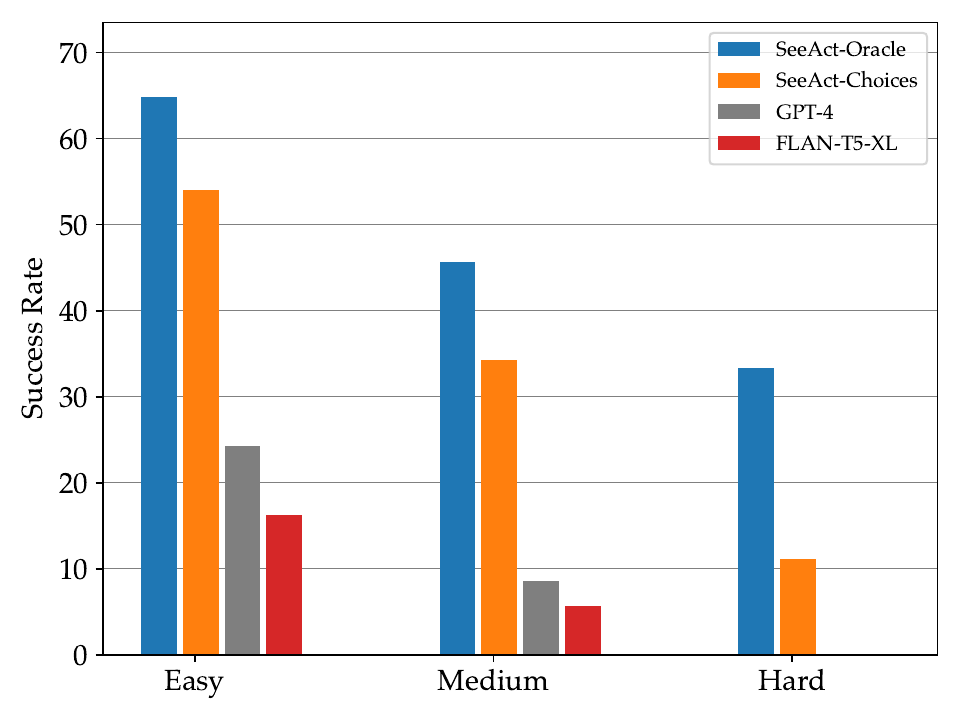}
    \caption{Whole task success rate across task difficulty levels. We categorize tasks based on the number of actions to complete, i.e., Easy: \num{1}-\num{4}, Medium: \num{5}-\num{9}, and Hard: \num{10}-\num{18}, with \num{37}, \num{35}, and \num{18} tasks in each group, respectively.}
    \label{fig:sr_difficulty}
\end{figure}

\noindent\textbf{Online Success Rate by Task Difficulty.}
We investigate the performance of web agents on tasks across different difficulty levels. We estimate the task difficulty based on the number of actions taken by annotators during action trace annotation. 
As shown in \autoref{fig:sr_difficulty}, the whole task success rate is negatively correlated with the number of actions---it decreases as the number of actions increases across all four methods. \lmmweb\textsubscript{Oracle} consistently outperforms other methods across all difficulty levels. 
Interestingly, the gap between \lmmweb\textsubscript{Oracle} and \lmmweb\textsubscript{Choice} enlarges on longer-horizon tasks.
This is understandable because grounding errors cascade to later steps; nonetheless, it further shows the challenge of grounding for GPT-4V and the need for better grounding methods.

\noindent\textbf{Error Analysis in Grounding via Image Annotation.} 
Set-of-mark prompting~\cite{Yang2023SetofMarkPU} uses a similar method as grounding via image annotation and has been shown effective on object- or scene-centric images~\cite{Lin2014MicrosoftCC, Plummer2015Flickr30kEC, Zhou2017ScenePT}. However, this grounding method is suboptimal on webpage screenshot images that are complex and contain rich semantic and spatial relationships.
To analyze the reasons behind the failures, we randomly sample \num{100} action predictions with correct action generation but wrong grounding results. We observes major types of errors as : (1) Making up bounding box \& label; (2) Failure to link bounding boxes with the correct labels. Illustrative examples are included in \autoref{sec:hallucination-examples}.

Our analysis reveals that \num{54}\% of the errors can be attributed to \gptv's tendency of visual illusion~\cite{Guan2023HallusionBenchAA}, where the model misinterprets and fabricates content over the image. Specifically, the target element described in action generation does not have a bounding box or a label on the bottom-left, where the model is supposed to generate "NA". However, the model falsely assumes the presence of a bounding box and makes up a label as the answer. Another \num{46}\% of errors are caused by \gptv's limitation in recognizing the relative position within an image. Specifically, the model is capable of identifying the target element within the bounding box. However, it struggles to correctly link the bounding box with its corresponding label.

\subsection{Case Study}
\gptv\ exhibits promising capabilities, ranging from speculative planning, webpage content reasoning, and error correction to surpassing the limitations of superficial textual similarity matching inherent in fine-tuned, text-only models.

\noindent\textbf{World Knowledge.}
\gptv\ demonstrates substantial advantages in tasks requiring certain knowledge over fine-tuned models at a smaller scale. As shown in \autoref{appendix_case_knowledge}, \gptv\ is able to identify the IATA code of the airport in Los Cabos as SJD. In contrast, smaller models are typically weaker at knowledge-intensive tasks and are also likely to lose knowledge during the fine-tuning process due to catastrophic forgetting. 

\noindent\textbf{World Model (for Websites).} 
\gptv\ exhibits the potential of a "world model" for websites. As shown in \autoref{appendix_case_planning}, \gptv\ can predict the state transitions on a website (e.g., what would happen if I clicked this button). Based on its awareness of website state transitions, \gptv\ can conduct speculative planning involving a sequence of subsequent actions in the future to complete the given task. 

\noindent\textbf{Error Correction Awareness.} 
\gptv\ also exhibits the awareness of error correction in the previous actions. In the example in \autoref{appendix_case_path}, it realizes that the mobile phone number is invalid due to the wrong format and generates the description of the action to correct this error. This highlights the model's potential for adaptation in online settings, where actions may not always follow pre-defined, ideal paths as in offline evaluations. This capability paves the way for adding robustness and reasonable dynamic planning. 

\section{Related Work}

\noindent\textbf{Web Agent.}
Many works have focused on improving web agents relying on the HTML document~\cite{deng2023mind2web,Gur2023ARW, Gur2022UnderstandingHW, Gur2023ARW, Kim2023LanguageMC, Sridhar2023HierarchicalPA}. However, a raw HTML document is often massive making it infeasible or cost-prohibitively to feed into LLMs directly. MindAct~\cite{deng2023mind2web} instead employs a small language model to rank each HTML element and selectively consider top elements as the context. WebAgent~\cite{Gur2023ARW} proposes an enhanced planning strategy by summarizing the HTML documents and decomposing the instruction into multiple sub-instructions. Another stream considers visual information for web agents~\cite{Shaw2023FromPT, Furuta2023MultimodalWN,Hong2023CogAgentAV}. Pix2Act~\cite{Shaw2023FromPT} leverages Pix2Struct~\cite{Lee2022Pix2StructSP} to parse screenshot images into simplified HTML to complete GUI-based tasks~\cite{Shaw2023FromPT, liu2018reinforcement, Shi2017WorldOB, Mazumder2020FLINAF, Yao2022WebShopTS}. WebGUM~\cite{Furuta2023MultimodalWN} and CogAgent~\cite{Hong2023CogAgentAV} pre-train an LMM with massive screenshot-HTML data to enhance its decision-making on real-world web navigation like Mind2Web.
While all these prior works show promise, generalizing to various web environments remains a challenge for existing models. Thus, \ourmethod\ explores recently released, more powerful LMMs such as GPT-4V and Gemini, to demonstrate their potential as generalist web agents with comprehensive online and offline evaluation and analysis. 
In a concurrent work~\cite {Yan2023GPT4VIW}, \gptv\ exhibits strong performance on mobile UI understanding, which is less complex than the desktop websites we study.



\noindent\textbf{Large Multimodal Models.}
\gptv~\cite{OpenAI2023GPT4TR} and Gemini~\cite{Anil2023GeminiAF} represent significant progress in LMMs. Several studies~\cite{Akter2023AnIL, OpenAI2023GPT4TR, Yang2023TheDO, Zhang2023GPT4VisionAA, Yang2023SetofMarkPU, Yan2023GPT4VIW} have highlighted their remarkable multimodal capabilities, emphasizing the advanced and versatile integration of visual and language reasoning abilities. Their performance on a series of benchmarks~\cite{Kazemzadeh2014ReferItGameRT, Goyal2016MakingTV, Hendrycks2020MeasuringMM, Saikh2022ScienceQAAN, Lu2022LearnTE, Zhong2023AGIEvalAH, Yue2023MMMUAM} also showcases remarkable capabilities on vision-and-language understanding and reasoning.
Although open-sourced models still exhibit a performance gap with \gptv, they have the advantages of controllability and ease of fine-tuning for various applications. For example, in CogAgent~\cite{Hong2023CogAgentAV}, LMMs are fine-tuned on HTML and screenshot image pairs to enhance webpage understanding ability and further enhanced with an image encoder for high-resolution image details. Ferret~\cite{You2023FerretRA} is finetuned to allow visual referring and grounding.

\noindent\textbf{Visual Grounding.} 
Despite LMMs having achieved remarkable vision-language understanding capabilities, they still face challenges in fine-grained visual grounding. Various visual prompting~\cite{Shtedritski2023WhatDC, Yang2023FineGrainedVP, Yang2023TheDO, Yan2023GPT4VIW} methods have been proposed to augment \gptv's image detail grounding ability by overlaying visual marks onto the image.
SoM~\cite{Yang2023SetofMarkPU} involves segmenting the image into semantically meaningful regions and overlaying an array of visual marks like numbers, alphabets, masks, or bounding boxes. 
Fine-tuning vision-language models with image-annotated data is effective. Kosmos-2~\cite{Peng2023Kosmos2GM} represents bounding box locations through textual location tokens. BuboGPT~\cite{Zhao2023BuboGPTEV} extract entities and find corresponding masks for objects in the image. Shikra~\cite{Chen2023ShikraUM} handles image detail referring and grounding by applying spatial coordinates as text tokens in inputs and outputs, respectively. Ferret~\cite{You2023FerretRA} represents regions with both discrete coordinates and continuous features along with a spatial-aware visual sampler to handle diverse spatial characteristics across various shapes.

\section{Conclusion}

In this work, we developed \ourmethod, a generalist web agent that harnesses the power of large multimodal models (LMMs) like GPT-4V to integrate visual understanding and acting on the web.
We showed that LMMs present a great promise for generalist web agents, with a success rate of \num{50}\% on live websites given an oracle grounding method. \gptv\ also exhibits impressive capabilities, such as error correction and speculative planning. However, fine-grained visual grounding is still a major challenge. The most effective grounding strategies we explored in this paper still exhibit a \num{20}-\num{25}\% performance gap compared to oracle grounding.
Future work should better leverage the unique properties of the Web, \emph{e.g.}, the known correspondence between HTML and visual elements, for improving grounding and reducing hallucinations from LMMs.
Furthermore, we show a significant discrepancy between online and offline evaluations, emphasizing the importance of online evaluation for an accurate assessment of a model’s capabilities. This discrepancy is largely due to the variability in potential plans for completing the same task, pointing to the dynamic nature of web interactions.

\section{Impact Statements}
\label{app:limitation}
Generalist web agents hold the potential to automate routine web tasks, enhance user experiences, and promote web accessibility, safety concerns related to their real-world deployment are also critical. These concerns span privacy issues, such as access to users' personal profiles, and sensitive operations, such as financial transactions or application form submissions. During the online evaluation, we noticed the possibility for these web agents to generate harmful actions on the web, and we manually validated the safety of all the actions before execution.
It is critical for further research to thoroughly assess and mitigate the safety risks associated with web agents, ensuring they are safeguarded against producing and executing harmful actions.
The code will also be released solely for research purposes, with the goal of making the web more accessible via language technologies under an OPEN-RAIL License. We are strongly against any potentially harmful use of the data or technology by any party.

\section*{Acknowledgments}
The authors would like to thank colleagues from the OSU NLP group for their thoughtful comments. This research was supported in part by ARL W911NF2220144 and Cisco.




\nocite{langley00}

\bibliography{main}
\bibliographystyle{icml2024}

\newpage
\appendix

\newpage
\onecolumn
\textbf{Table of Content:}
\begin{enumerate}[nosep]
    \item{\cref{appendix:methods}: Offline Experiments Method Details}
    \item{\cref{appendix:markup-oblation}: Markup Type Ablation Study}
    \item{\cref{{appendix:online-exp-details}}: Online Experiment Details}
    \item{\cref{appendix:action-generation}:Offline Experiment Prompts}
        
    \item{\cref{sec:hallucination-examples}: Error Examples for Grounding via Image Annotation}
    \item{\cref{appendix_case_planning}: Strong Capability of Planning}
    \item{\cref{appendix_case_identical}: Challenges in Grounding via Textual Choices}
    \item{\cref{appendix_case_knowledge}: Knowledge and Reasoning Requirements}
    \item{\cref{appendix_case_path}: Path Variation and Awareness of Error Correction}
\end{enumerate}
\newpage

\section{Offline Experiments Method Details}
\label{appendix:methods}

\noindent \textbf{FLAN-T5.} 
We fine-tune FLAN-T5 using a left-to-right language modeling objective with the target sequence of ground-truth actions in the Mind2Web training data. The fine-tuned FLAN-T5 then serves as the backbone for inference, enabling action generation in the target format for parsing. 

\noindent \textbf{BLIP-2-T5.} 
The BLIP-2 model combines a vision encoder and an LLM with a bridging component for modality connection. We jointly fine-tune the LLM and the bridge module on \dataset\ training data while keeping the vision encoder frozen. For the vision encoder, we leverage the ViT-L/14 pre-trained from CLIP~\cite{clip} with an image resolution of \num{2048}. To ensure a fair comparison with the FLAN-T5-based text-only model, we choose FLAN-T5 as the language model and initialize it with the parameters fine-tuned on \dataset.

\noindent \textbf{GPT-3.5 and GPT-4.} 
We also conduct experiments with text-only LLMs, specifically GPT-3.5-turbo-0613 and GPT-4-turbo-1106-preview, using in-context learning in 3-shot settings. We use the same multiple-choice formulation and include three demonstration examples for in-context learning as specified in MindAct. 

\noindent \textbf{SeeAct} 
We experiment with GPT-4-vision-preview, Gemini Pro Vision, LLaVA-1.5. 
Gemini Pro Vision supports only single-turn conversations; therefore, we merge the two turns used in other models for compatibility. 

\noindent \textbf{CogAgent} 
We utilize the cogagent-chat-hf checkpoint that hasn't been fine-tuned on Mind2Web for experiments. 

\section{Markup Type Ablation Study}
\label{appendix:markup-oblation}
The markup types might influence model performance as shown in~\citet{Yang2023SetofMarkPU, Yan2023GPT4VIW, Koh2024VisualWebArenaEM}.
We first tested grounding via image annotation through different types of text labels of numerical value, single-digit characters, and two-digit characters, at two different positions to choose a relatively better markup type. The results are shown in \autoref{tab:image_annotation_oblation}. 

\begin{table}[h]
\caption{ Grounding via Image Annotation with different markup types and locations. Method with * mark means the annotated image is used in action generation.
}
\centering
\small
\tabcolsep 3.5pt
\renewcommand\arraystretch{1.0}
\begin{tabular}{lllccc}
\toprule
&Label&Location  & Ele. Acc & Op. F1 & Step SR \\
\midrule
\cmidrule(r){1-3}

&\multirow{2}{*}{Number}&
Bottom-Left&\bfseries\num{27.0}&\num{73.7}&\bfseries\num{24.3}\\
&&Bottom-Center&\num{23.0}&\num{76.4}&\num{21.8}\\
\addlinespace[0.1em]\hdashline\addlinespace[0.1em]

&\multirow{2}{*}{Single Letter}&Bottom-Left&\num{19.4}&\bfseries\num{81.0}&\num{17.2}\\
&&Bottom-Center&\num{19.7}&\num{78.8}&\num{19.7}\\
\addlinespace[0.1em]\hdashline\addlinespace[0.1em]

&\multirow{2}{*}{Double Letter}& Bottom-Left&\num{19.8}&\num{68.3}&\num{18.3}\\
&&Bottom-Center&\num{22.4}&\num{74.6}&\num{22.4}\\

\addlinespace[0.1em]\hdashline\addlinespace[0.1em]
&\textsc{Number*}&Bottom-Left*&\num{26.6}&\num{73.9}&\num{22.3}\\
\bottomrule
\end{tabular}
\vspace{-5pt}
\label{tab:image_annotation_oblation}
\end{table}

\section{Online Experiment Details}
\label{appendix:online-exp-details}
We develop an online evaluation tool using Playwright to load webpages, acquire textual representation of interactive elements, and perform operations generated by web agents. We manually monitor each step of the model and assess whether it finishes the tasks. Attempts to log in, make final submissions, or perform other potentially harmful actions are prohibited to avoid negative consequences.

\noindent\textbf{MindAct.} We adhere strictly to the original settings in MindAct-FLAN-T5 and MindAct-GPT-4, as in offline experiment. For consistency with the MindAct framework and models, we use the scripts from Mind2Web for processing webpages, elements, and generating options, employing the same action space: \texttt{Click}, \texttt{Type}, and \texttt{Select}.

\noindent\textbf{\lmmweb\textsubscript{Oracle}.} In the oracle setting, we manually implement the model's intended actions. The action history is automatically generated by the model, with an added requirement to summarize actions in the "Element", "Operation", "Value" format in another turn conversation. To avoid overly long screenshots, we use screenshots of current views, and hence allow intentions of scrolling.
  
\noindent\textbf{\lmmweb\textsubscript{Choice}.} For a relatively fair comparison, we still adopt the ranker and top-\num{50} candidate setting, batching them into three option groups as described in offline experiments. We enable \texttt{PRESS ENTER} and \texttt{TERMINATE} for the model to make confirmation or stop the process.

During our tests, pop-up ads on webpages were manually closed. The MindAct model, not trained on handling pop-up ads, lacks the feature to automatically manage them, potentially causing stalls. In contrast, \lmmweb\ models can proactively suggest closing ads through visual analysis and reasoning.

\section{Offline Experiment Prompts}
\label{appendix:action-generation}
\begin{figure*}[h]
    \centering
    \includegraphics[width=0.9\linewidth]{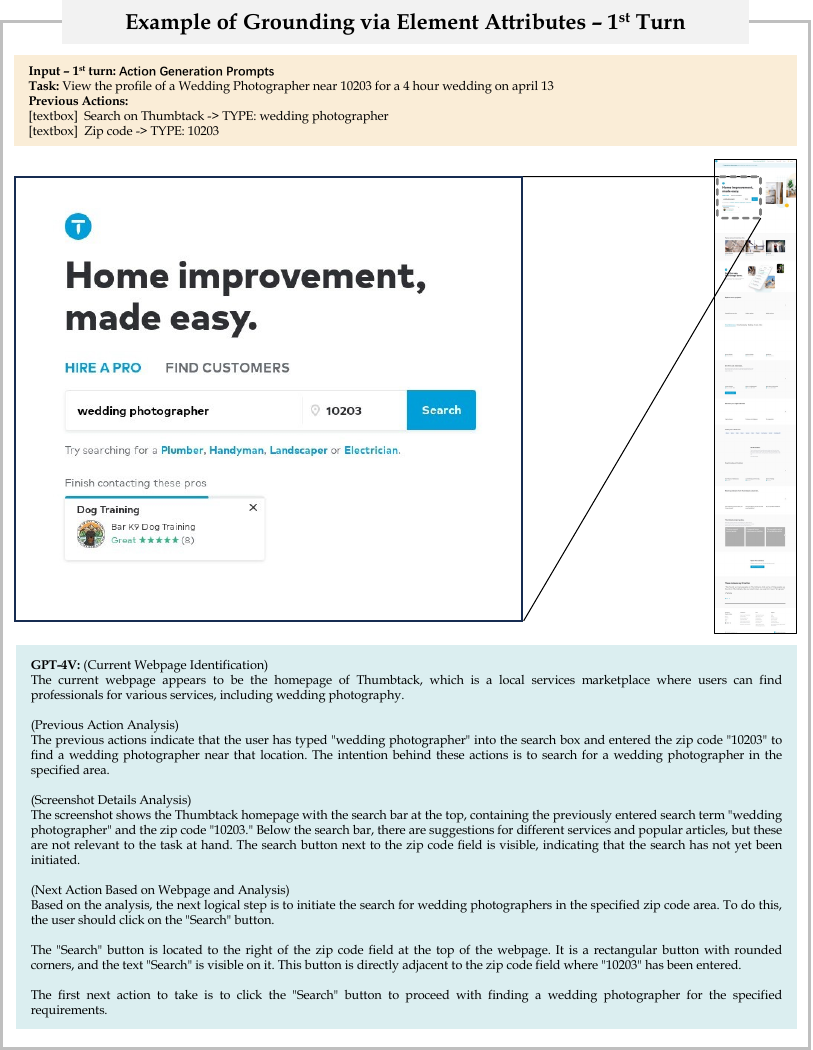}
    \caption{An example of action generation in grounding via element attributes.}
    \label{fig:example-att-1}
\end{figure*}

\begin{figure*}[h]
    \centering
    \includegraphics[width=0.9\linewidth]{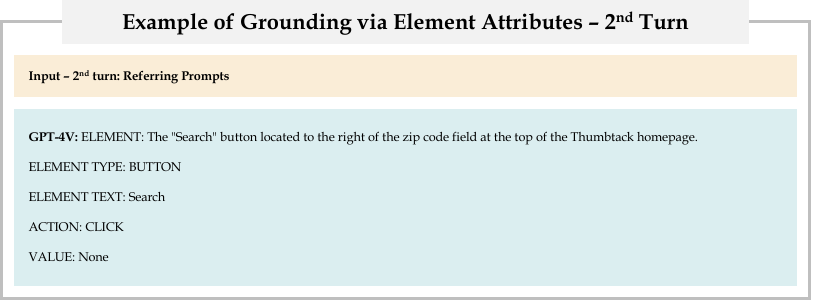}
    \caption{An example of grounding via element attributes after action generation.}
    \label{fig:example-att-2}
\end{figure*}

\begin{figure*}[h]
    \centering
    \includegraphics[width=0.9\linewidth]{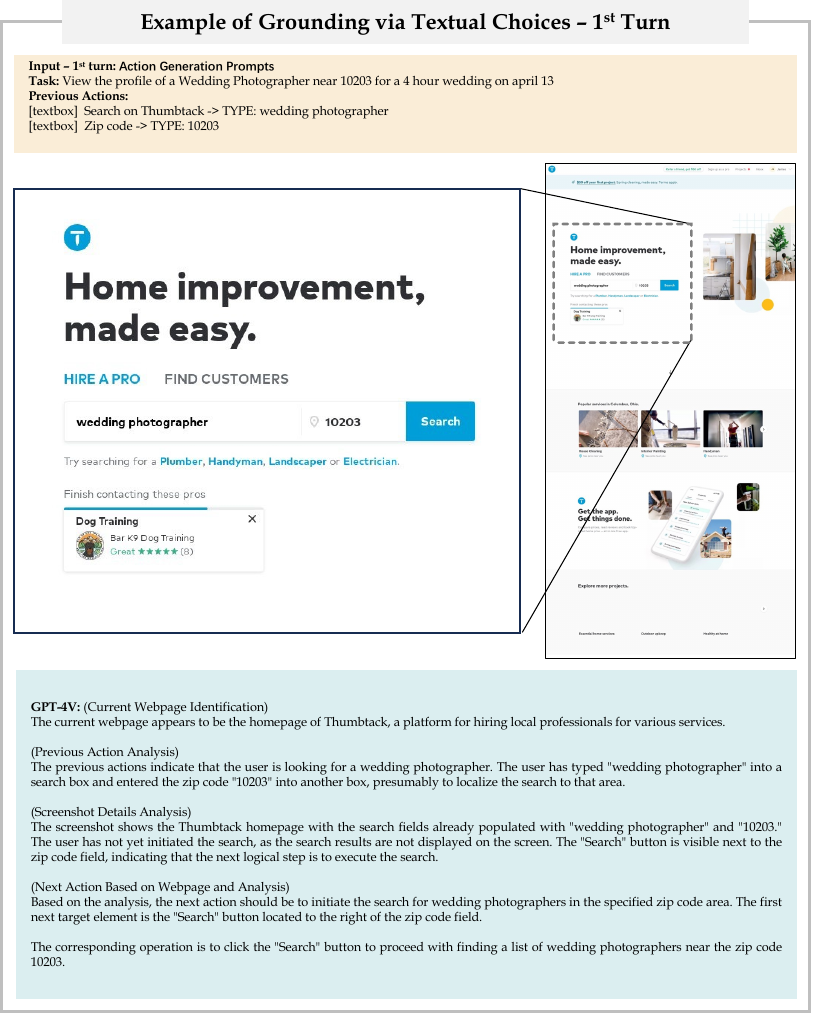}
    \caption{An example of action generation in grounding via textual choices.}
    \label{fig:example-text-choice-1}
\end{figure*}

\begin{figure*}[h]
    \centering
    \includegraphics[width=0.9\linewidth]{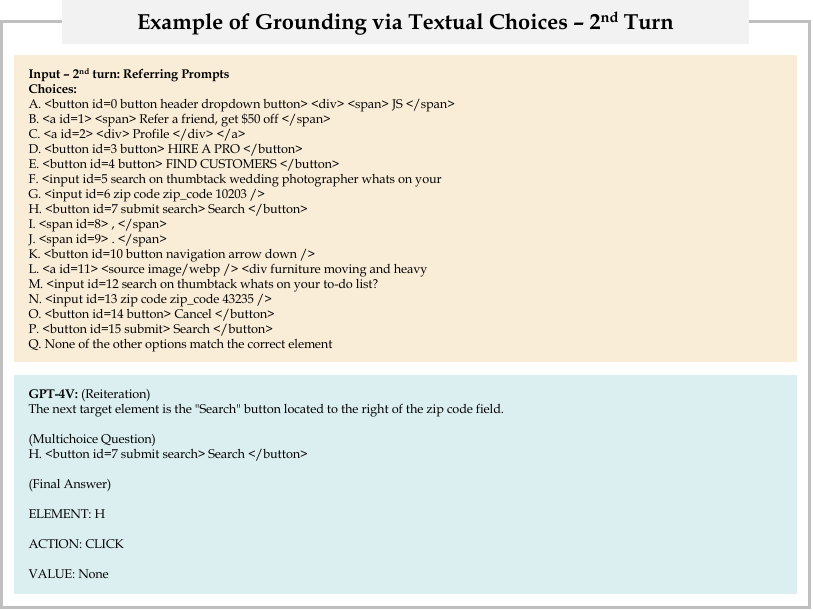}
    \caption{An example of grounding via textual choices after action generation.}
    \label{fig:example-text-choice-2}
\end{figure*}

\begin{figure*}[h]
    \centering
    \includegraphics[width=0.9\linewidth]{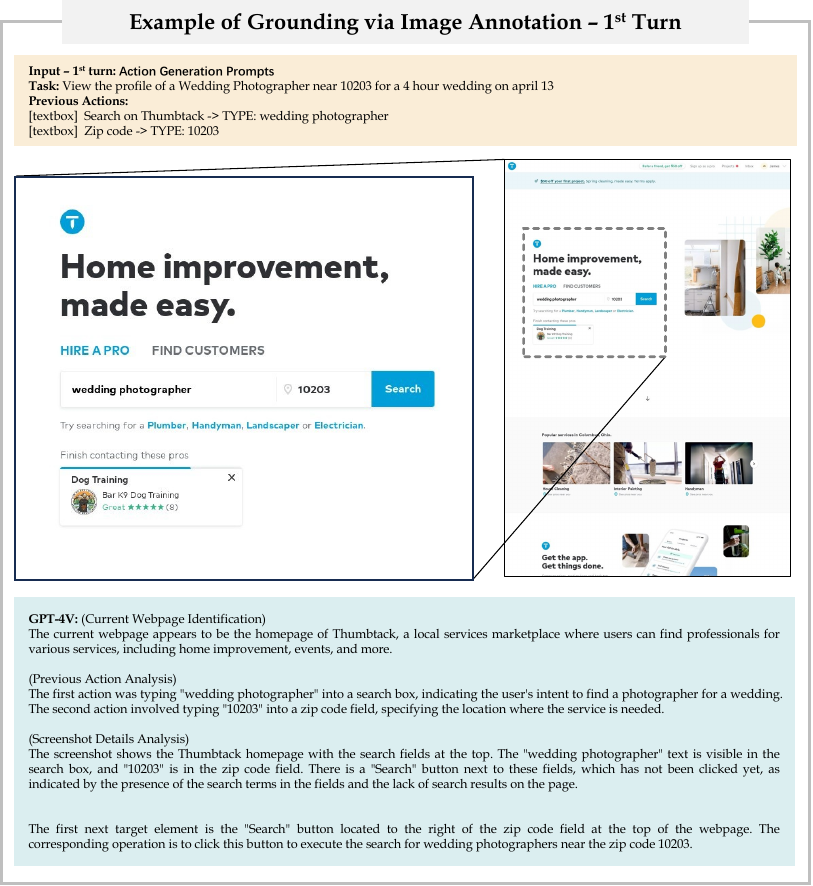}
    \caption{An example of grounding via image annotation.}
    \label{fig:example-exp2}
\end{figure*}

The prompt for action generation is shown in \autoref{tab:task_generation_prompt}. For grounding via textual choices, image annotation, and element attributes, the prompts are shown in \cref{tab:exp3,tab:exp4,tab:exp2}, along with specific tasks and examples in \cref{fig:example-att-1,fig:example-att-2,fig:example-text-choice-1,fig:example-text-choice-2,fig:example-exp2}.
\begin{table*}[h]
    \centering
    \small
    \caption{Prompt for \lmmweb\ Action Generation with LMMs.}
    \begin{tabular}{lp{10cm}}
    \toprule
        \textbf{System Role} & Imagine that you are imitating humans doing web navigation for a task step by step. At each stage, you can see the webpage like humans by a screenshot and know the previous actions before the current step decided by yourself through recorded history. You need to decide on the first following action to take. You can click an element with the mouse, select an option, or type text with the keyboard. (For your understanding, they are like the click(), select\_option() and type() functions in playwright respectively) One next step means one operation within the three.\\
        \bottomrule
        \textbf{Action Generation} & You are asked to complete the following task: \{TASK\}
        
        \\

        &Previous Actions: 
        
        \{PREVIOUS ACTIONS\}

        \\
        
        &The screenshot below shows the webpage you see. Follow the following guidance to think step by step before outlining the next action step at the current stage:

        \\
        
        &(Current Webpage Identification)
        
Firstly, think about what the current webpage is.

        \\
        
        &(Previous Action Analysis)
        
Secondly, combined with the screenshot, analyze each step of the previous action history and their intention one by one. Particularly, pay more attention to the last step, which may be more related to what you should do now as the next step.

        \\
        
        &(Screenshot Details Analysis)
        
Closely examine the screenshot to check the status of every part of the webpage to understand what you can operate with and what has been set or completed. You should closely examine the screenshot details to see what steps have been completed by previous actions even though you are given the textual previous actions. Because the textual history may not clearly and sufficiently record some effects of previous actions, you should closely evaluate the status of every part of the webpage to understand what you have done.

        \\
        
        &(Next Action Based on Webpage and Analysis)
        
Then, based on your analysis, in conjunction with human web browsing habits and the logic of web design, decide on the following action. And clearly outline which element in the webpage users will operate with as the first next target element, its detailed location, and the corresponding operation.

\\
        
        &To be successful, it is important to follow the following rules: 

1. You should only issue a valid action given the current observation. 

2. You should only issue one action at a time.\\
        \bottomrule
    \end{tabular}
    \label{tab:task_generation_prompt}
\end{table*}
\newpage

\begin{table*}[h]
    \centering
    \small
    \caption{Prompt for \lmmweb\ grounding via element attributes. We make a slight modification to enhance action generation and only show the modified part here to save space, as well as the prompts in \autoref{tab:exp4} and \autoref{tab:exp2}.
    }
    \begin{tabular}{lp{10cm}}
    \toprule
 \textbf{System Role} & Same as \autoref{tab:task_generation_prompt}\\
        \bottomrule
        \textbf{Action Generation} &  Slightly modified from  \autoref{tab:task_generation_prompt}\\

&...\\

        &(Next Action Based on Webpage and Analysis)

Then, based on your analysis, in conjunction with human web browsing habits and the logic of web design, decide on the following action. And clearly outline which element in the webpage users will operate with as the first next target element, its detailed location, and the corresponding operation. Please also closely examine the screenshot to adequately describe its position relative to nearby elements and its textual or visual content (if it has). If you find multiple elements similar to your target element, use a more precise description to ensure people can distinguish your target element from them through your answer.\\
        &...\\

         \bottomrule 
        \textbf{Format Answer} & (Final Answer)
Finally, conclude your answer using the format below. Ensure your answer is strictly adhering to the format provided below. Please do not leave any explanation in your answers of the final standardized format part, and this final part should be clear and certain. The element, element type, element text, action and value should be in five separate lines.

\\
        
        &Format:

\\
        
        &ELEMENT: Please describe which element you need to operate with. Describe it as detailed as possible, including what it is and where it is.

\\
        
        &ELEMENT TYPE: Please specify its type from these options: BUTTON, TEXTBOX, SELECTBOX, or LINK.

\\
        
        &ELEMENT TEXT: Please provide the exact text displayed on the element. Do not invent or modify the text; reproduce it as-is from the screenshot.

\\
        
        &ACTION: Choose an action from \{CLICK, TYPE, SELECT\}.

\\
        
        &VALUE: Provide additional input based on ACTION.

\\
        
        &The VALUE means:
If ACTION == TYPE, specify the text to be typed.
If ACTION == SELECT, specify the option to be chosen.
If ACTION == CLICK, write "None".\\
        \bottomrule
    \end{tabular}
    \label{tab:exp3}
\end{table*}
\newpage

\begin{table*}[h]
    \centering
    \small
    \caption{Prompt for \lmmweb\ grounding via textual choices. }
    \begin{tabular}{lp{10cm}}
    \toprule
     \textbf{System Role} & Same as \autoref{tab:task_generation_prompt}\\
        \bottomrule
        \textbf{Action Generation} &  Same as  \autoref{tab:task_generation_prompt}\\
         \bottomrule 
        \textbf{Referring Description} & (Reiteration)
        
First, reiterate your next target element, its detailed location, and the corresponding operation.

\\
        
        &(Multichoice Question)

Below is a multi-choice question, where the choices are elements in the webpage. From the screenshot, find out where and what each one is on the webpage. Then, determine whether one matches your target element. Please examine the choices one by one. Choose the matching one. If multiple options match your answer, choose the most likely one by re-examining the screenshot, the choices, and your further reasoning.

\\
        
        &If none of these elements match your target element, please select [None of the other options match the correct element].

        A. [CHOICE A]

        B. [CHOICE B]

        ...

        \\
         \bottomrule 
        \textbf{Format Answer} & (Final Answer)
        
Finally, conclude your answer using the format below. Ensure your answer is strictly adhering to the format provided below. Please do not leave any explanation in your answers of the final standardized format part, and this final part should be clear and certain. The element choice, action, and value should be in three separate lines.

\\
        
        &Format:

\\
        
        &ELEMENT: The uppercase letter of your choice.

\\
        
        &ACTION: Choose an action from \{CLICK, TYPE, SELECT\}.

\\
        
        &VALUE: Provide additional input based on ACTION.

\\
        
        &The VALUE means:
If ACTION == TYPE, specify the text to be typed.
If ACTION == SELECT, specify the option to be chosen.
If ACTION == CLICK, write "None".\\
        \bottomrule
    \end{tabular}
    \label{tab:exp4}
\end{table*}
\newpage

\begin{table*}[h]
    \centering
    \small
    \caption{
    Prompt for \lmmweb\ grounding via image annotation.}
    \begin{tabular}{lp{10cm}}
    \toprule
     \textbf{System Role} & Same as \autoref{tab:task_generation_prompt}\\
        \bottomrule
        \textbf{Action Generation} &  Same as \autoref{tab:task_generation_prompt}\\
         \bottomrule 
        \textbf{Referring Description} &(Reiteration)
        
First, reiterate your next target element, its detailed location, and the corresponding operation.

\\
        
        &(Verification with the Screenshot)
        
Then, please closely re-examine the screenshot to find whether your target element is marked by a red bounding box and has a white number on a black background at the bottom left corner of the bounding box, which is positioned closely next to the bounding box. If yes, use that number for your final answer. If not, please do not make them up. If it is not marked, please output "NA" as your target element in the following final answer part.\\
         \bottomrule 
        \textbf{Format Answer} & (Final Answer)
        
Finally, conclude your answer using the format below. Ensure your answer is strictly adhering to the format provided below. Please do not leave any explanation in your answers of the final standardized format part, and this final part should be clear and certain. The element choice, action, and value should be in three separate lines.

\\
        
        &Format:

\\
        
        &ELEMENT: The number of your choice.

\\
        
        &ACTION: Choose an action from \{CLICK, TYPE, SELECT\}.

\\
        
        &VALUE: Provide additional input based on ACTION.

\\
        
        &The VALUE means:
If ACTION == TYPE, specify the text to be typed.
If ACTION == SELECT, specify the option to be chosen.
If ACTION == CLICK, write "None".\\
        \bottomrule
    \end{tabular}
    \label{tab:exp2}
\end{table*}
\newpage

\newpage

\newpage

\newpage

\newpage

\newpage

\begin{figure*}[h]
    \centering
    \includegraphics[width=0.9\linewidth]{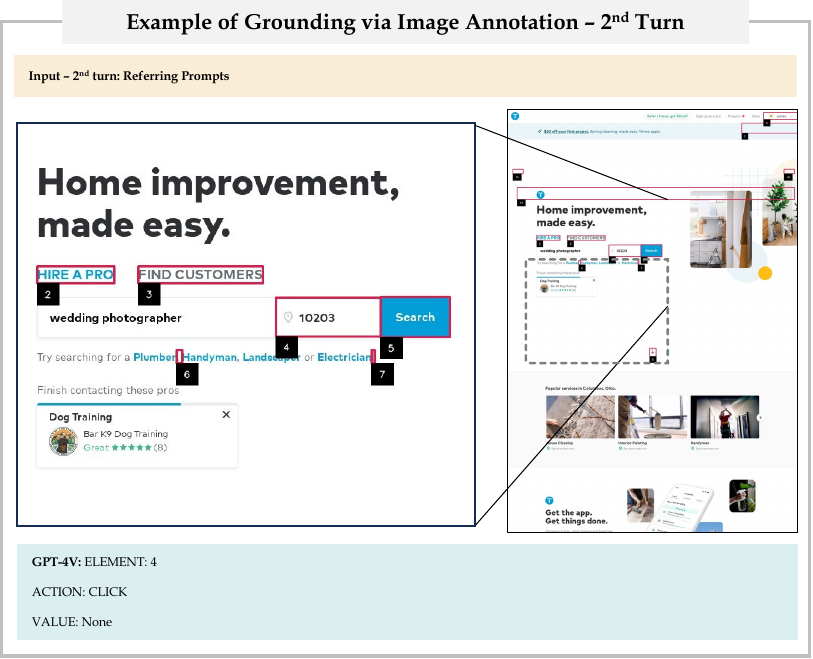}
    \caption{An example of grounding via image annotation.}
    \label{fig:example-exp2_2nd}
\end{figure*}
\newpage

\section{Error Examples for Grounding via Image Annotation}\label{sec:hallucination-examples}

\begin{figure*}[h]
    \centering
    \includegraphics[width=0.7\linewidth]{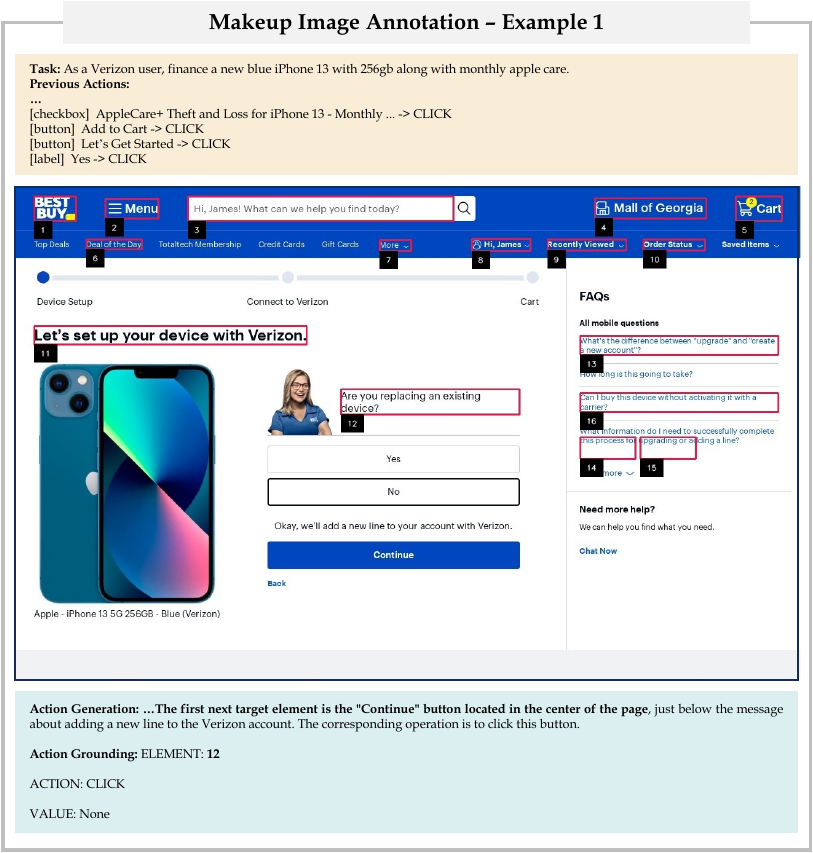}
    \caption{
    In this example, the model describes a correct element in action generation. However, the identified element is absent from the set of candidate elements. Despite this, the model erroneously assigns it the index number "12".    
    }
    \label{fig:makeup1}
\end{figure*}

\begin{figure*}[h]
    \centering
    \includegraphics[width=0.7\linewidth]{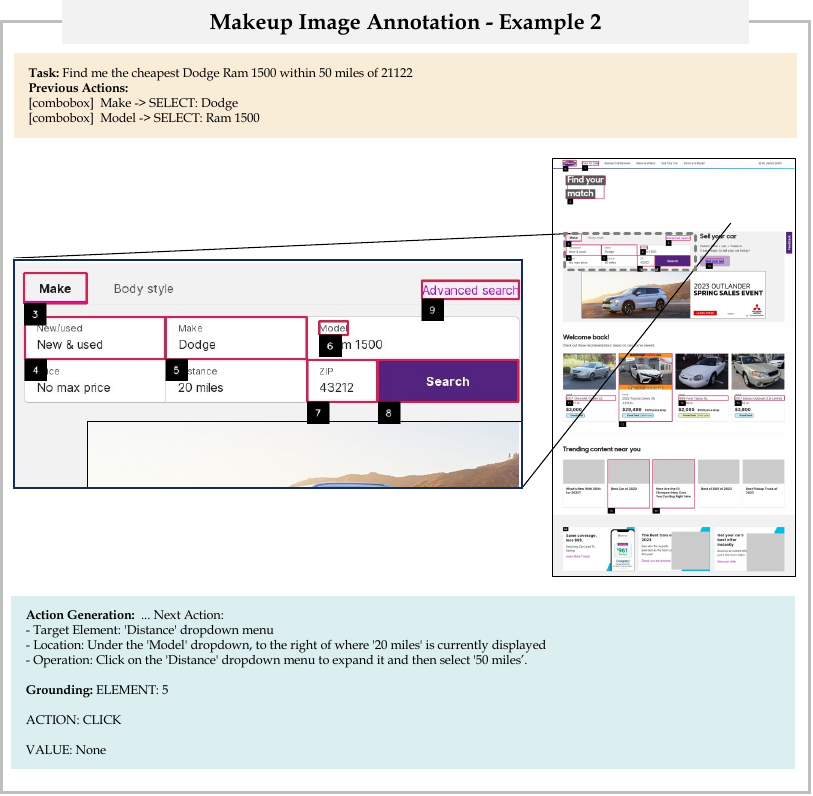}
    \caption{In this example, the model predicts the appropriate element. Nevertheless, the identified element is not present in the provided image options. Despite this, the model erroneously assigns it with the label '5' nearby. }
    \label{fig:makeup2}
\end{figure*}

In the method of grounding via image annotation, we observe significant hallucination errors that can be classified into the following categories:

\noindent\textbf{Making up bounding box \& label.} 
In our grounding method, if the correct element is absent from the set of candidate elements, the model is anticipated to generate "NA" as the answer. However, as depicted in \autoref{fig:makeup1} and \autoref{fig:makeup2}, the model erroneously claims the element is included within a red bounding box and makes up a wrong index label as the answer.

\begin{figure*}[h]
    \centering
    \includegraphics[width=0.9\linewidth]{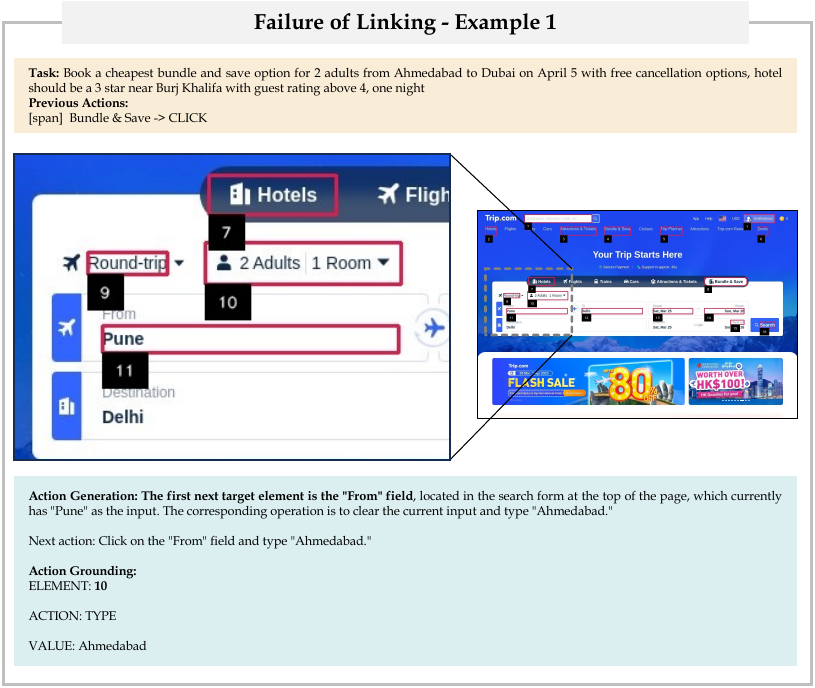}
    \caption{In this case, while the model predicts the appropriate element, it incorrectly associates the element with the nearby label '10' instead of the correct label '11'.}
    \label{fig:link1}
\end{figure*}

\begin{figure*}[h]
    \centering
    \includegraphics[width=0.9\linewidth]{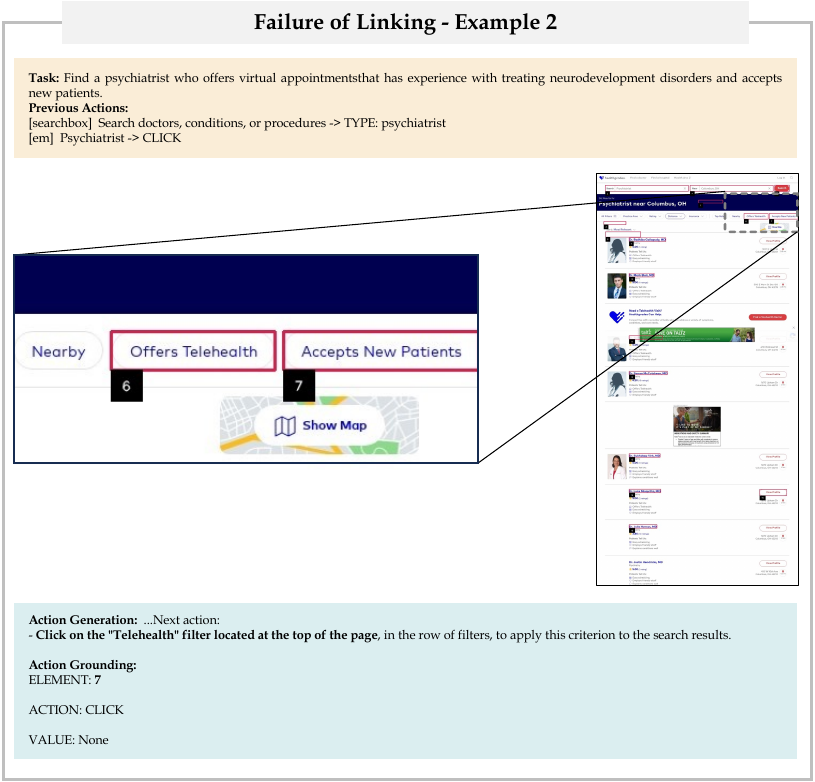}
    \caption{In this case, while the model predicts the appropriate element, it incorrectly associates the element with the nearby label '7' instead of the correct label '6'.}
    \label{fig:link2}
\end{figure*}

\noindent\textbf{Failure to link bounding boxes with the correct labels.}
Another challenge arises in accurately linking bounding boxes to their corresponding index labels. This challenge can be attributed to both LMMs' limitations in understanding relative spatial positions and the complex, dense layout of webpage elements. The model often mistakenly associates the labels of adjacent elements (as illustrated in \autoref{fig:link1} and \autoref{fig:link2}), rather than accurately predicting the intended index label for the targeted element.

\section{Strong Capability of Planning}\label{appendix_case_planning}

\begin{figure*}[h]
    \centering
    \includegraphics[width=0.8\linewidth]{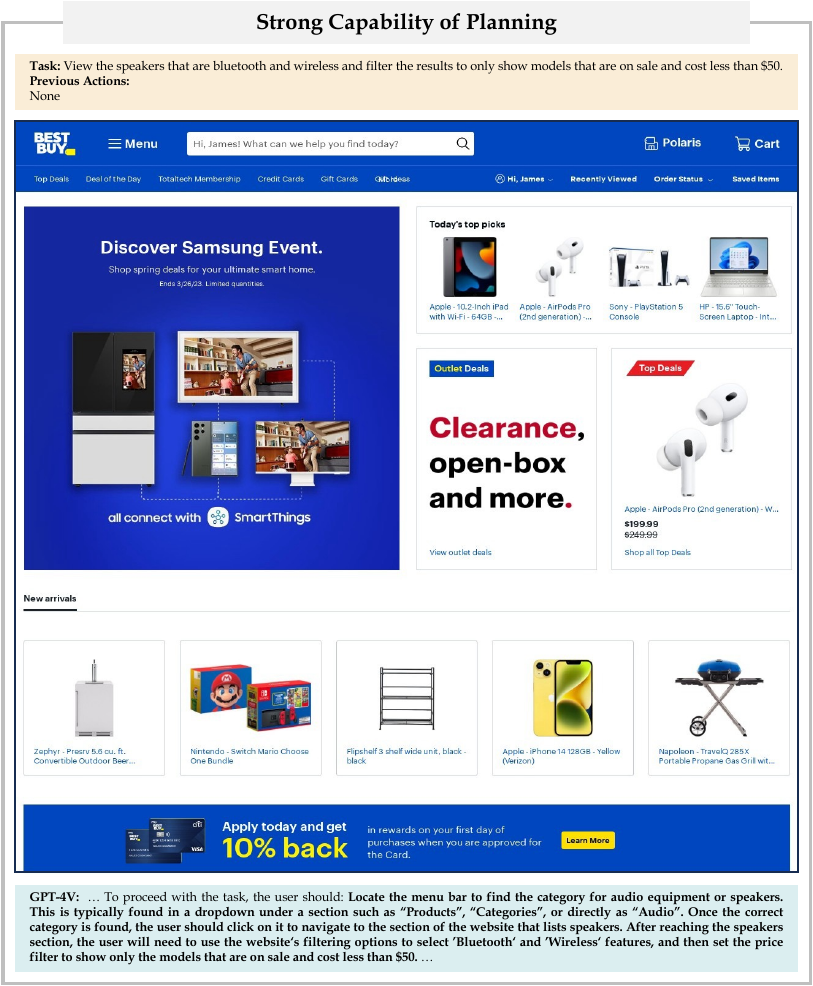}
    \caption{In this example, the model generates a comprehensive plan for the task, including subsequent actions on following pages that are not currently visible.}
     \label{fig:planning}
\end{figure*}

\begin{figure*}[h]
    \centering
    \includegraphics[width=0.9\linewidth]{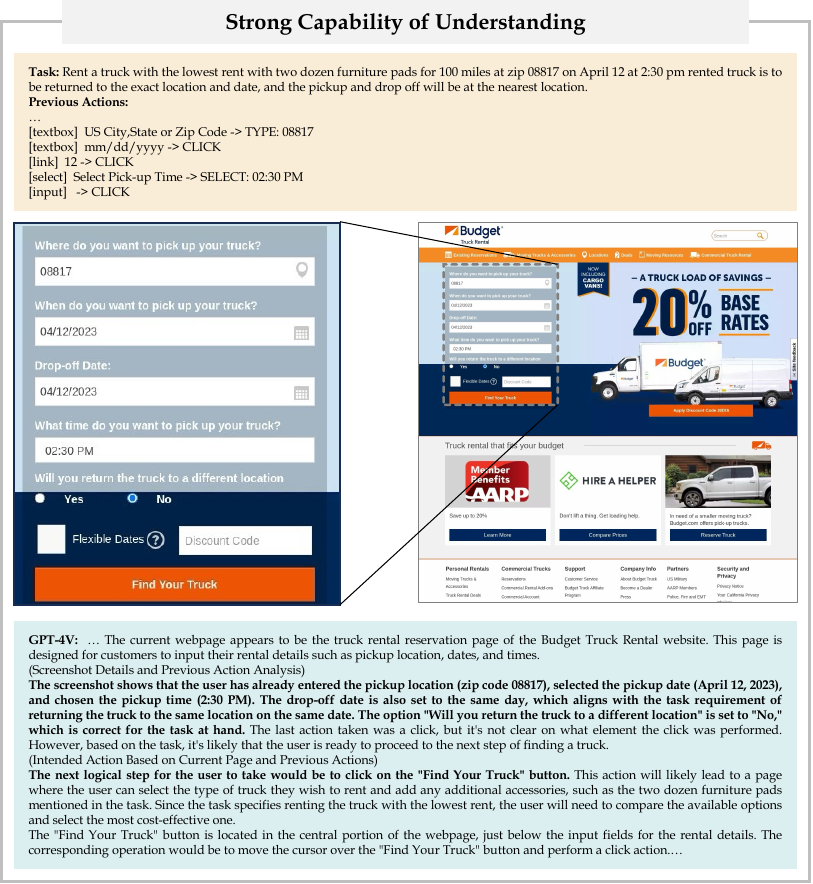}
    \caption{
    In this case, two critical pieces of information are inadequately captured by the textual history. Firstly, the website automatically set the drop-off date to the same day. Secondly, secondly, the 'No' button was selected (However it was missed in previous actions due to the button's lack of text). Nevertheless, the model discerns these details through a meticulous analysis of the screenshot, enabling it to make the correct decision for the next step.
    }
    \label{fig:understanding}
\end{figure*}

\gptv\ shows remarkable understanding and planning capabilities during our experiments. As depicted in \autoref{fig:planning}, the model is capable of understanding the website and generating a full plan for the given task involving multiple low-level tasks. Specifically, \gptv\ could understand reasonably well about the process and the remaining work of the task by its careful examination of the webpage, as shown in \autoref{fig:understanding}.

\section{Challenges in Grounding via Textual Choices}\label{appendix_case_identical}
\begin{figure*}[h]
    \centering
    \includegraphics[width=0.9\linewidth]{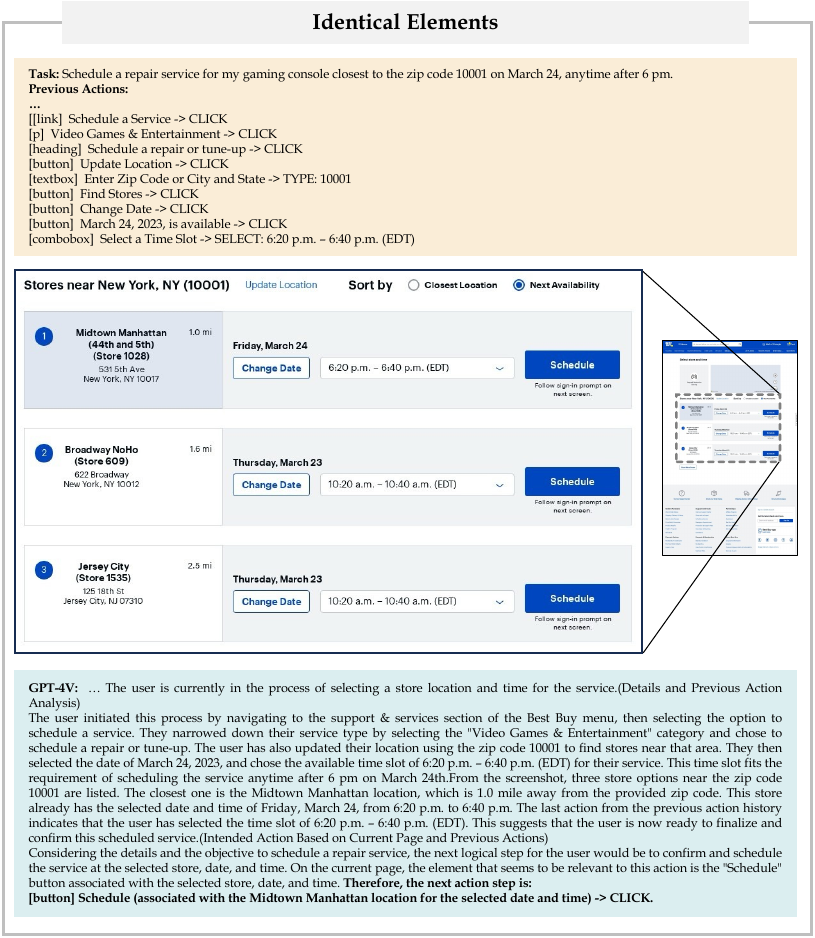}
    \caption{In this example, there are three identical 'Schedule' buttons, making it impossible for \lmmweb\textsubscript{Choice} to distinguish among them. We empirically find that the model tends to choose the first one among the choices.}
    \label{fig:identical}
\end{figure*}

Although textual choices achieved the best results among the three grounding approaches, it still suffers from challenges of similar or identical elements which are common in webpages. The model tends to choose the first text choice that seemingly corresponds to its intention. Moreover, this is inevitable, as web pages indeed contain many elements that may even have exact identical HTML information, as the "Schedule" button shown in \autoref{fig:identical}.

\section{Knowledge and Reasoning Requirements}\label{appendix_case_knowledge}
\begin{figure*}[h]
    \centering
    \includegraphics[width=0.9\linewidth]{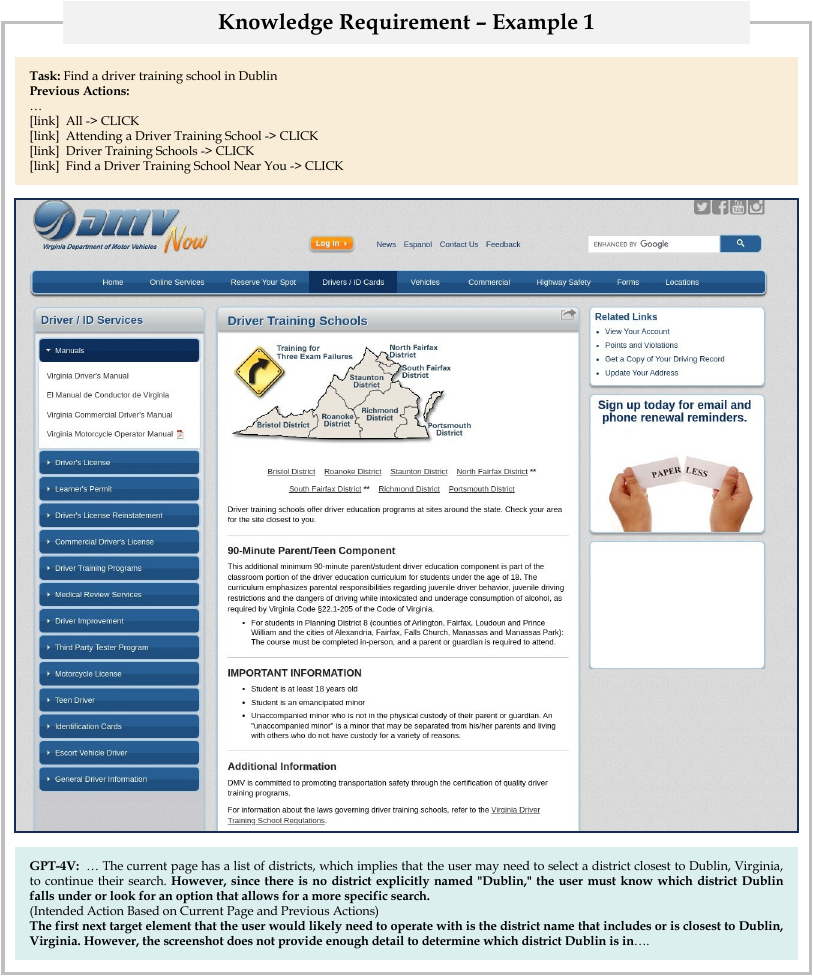}
    \caption{In this example, the task necessitates knowledge about which district Dublin is located in.}
    \label{fig:knowledge_1}
\end{figure*}

\begin{figure*}[h]
    \centering
    \includegraphics[width=0.9\linewidth]{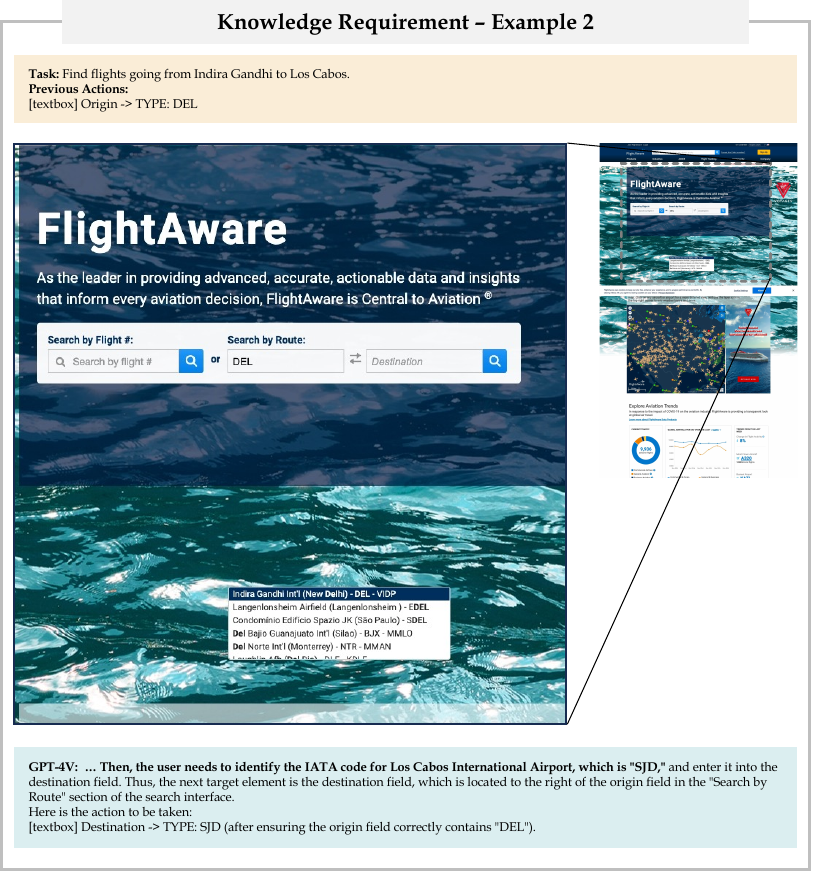}
    \caption{In this example, the task requires knowledge of the IATA code for Los Cabos International Airport. \gptv\ accurately provides the correct code.}
    \label{fig:knowledge_2}
\end{figure*}

Some tasks require a certain degree of reasoning and knowledge, which may be challenging for fine-tuned models like MindAct. For instance, the task in \autoref{fig:knowledge_1} necessitates the model to know the specific district of Dublin in Virginia. In the task of \autoref{fig:knowledge_2}, the model correctly provided the IATA airport code of airports in Indira Gandhi and Los Cabos.

\section{Path Variation and Awareness of Error Correction}\label{appendix_case_path}
\begin{figure*}[h]
    \centering
    \includegraphics[width=0.8\linewidth]{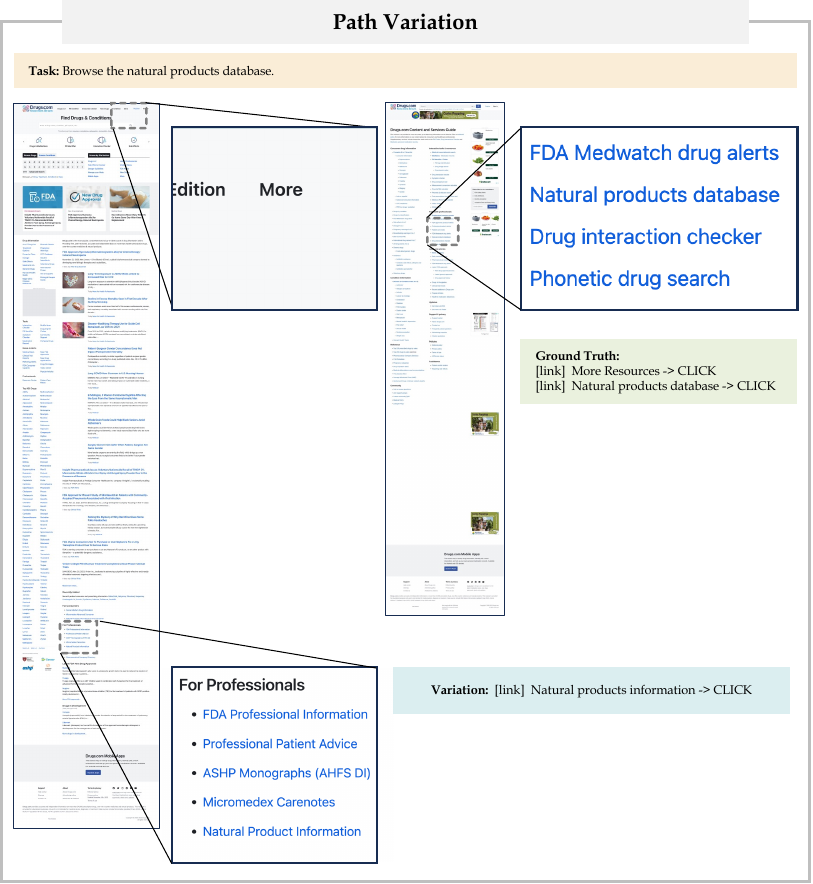}
    \caption{In this example, the ground truth in Mind2Web is to firstly click "More", then click "Natural products database" on the second page. In contrast, the model identifies a more direct approach, achieving the target page through 'Natural products information' on the first page.}
    \label{case:path}
\end{figure*}

On webpages, multiple paths often exist to accomplish a given task. For instance, varying the execution order of actions within an interchangeable sequence can result in diverse routes to task completion. Additionally, the agent can navigate to different webpages but still accomplish the give tasks. \autoref{case:path} presents a straightforward example where the model chose a more direct route that differs from the ground truth annotated in the dataset. 

\begin{figure*}[h]
    \centering
    \includegraphics[width=0.9\linewidth]{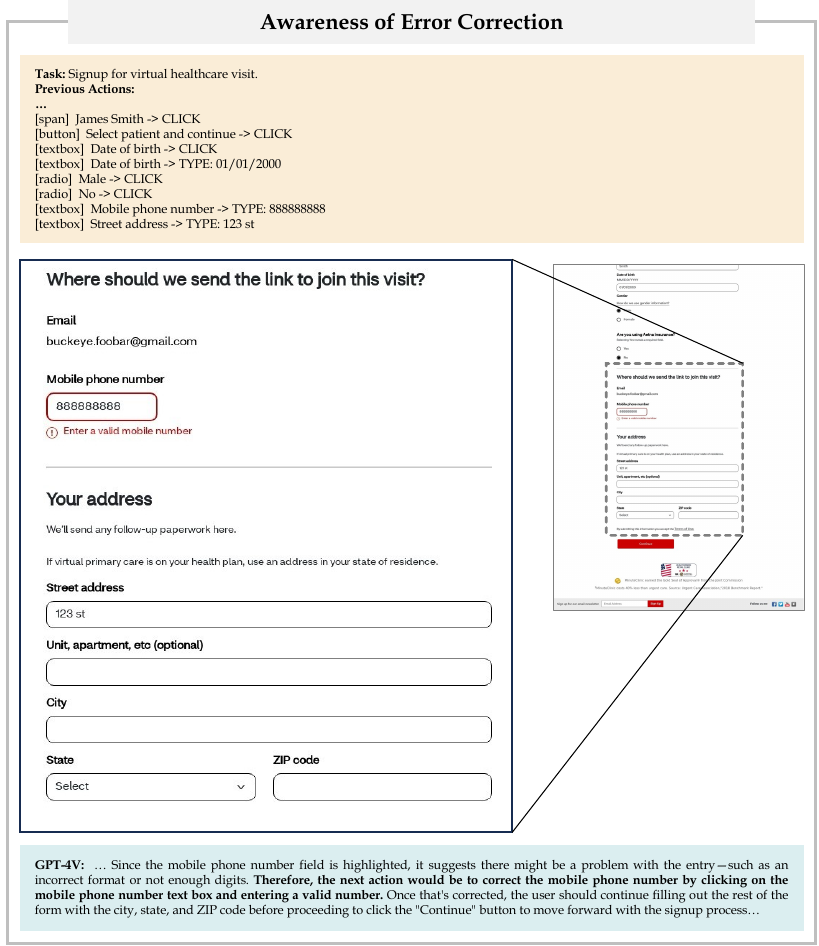}
    \caption{In this example, the webpage displays an error message indicating an invalid phone number, a consequence of prior actions. The model identifies this error and prioritizes its immediate rectification, foregoing the subsequent planned steps.}
    \label{fig:case-error-correction}
\end{figure*}

When running on live website, the agent's previous action histories is likely to be filled with redundant, unnecessary, erroneous, failed operations generated, or merely exploratory attempts by the model, resulting in a final path that deviates significantly from the ground truth. Despite these circumstances, the model can still accomplish the task amidst numerous incorrect explorations. The process of exploration and correction requires the model to possess a sense of self-correction. As shown in \autoref{fig:case-error-correction}, \gptv\ demonstrates this awareness of correcting errors caused by previous steps.

\newpage

\end{document}